%% file: main.tex
\pdfoutput=1
\documentclass[format=acmsmall, review=false, screen=true]{acmart}
\usepackage{multirow}
\usepackage{algorithmic}
\usepackage{amsthm,amsmath}
\usepackage{caption}
\usepackage{xcolor}
\usepackage[ruled,linesnumbered]{algorithm2e}
\usepackage{enumitem}
\usepackage{csquotes}
\usepackage{balance}
\usepackage{multicol}
\usepackage{makecell}
\usepackage{marvosym}
\usepackage{fancybox}
\usepackage{soul}
\usepackage{listings}
\usepackage{graphicx}
\usepackage{subcaption}
\usepackage{tikz}
\usepackage{booktabs}
\usepackage{wrapfig}
\usepackage[T1]{fontenc}
\definecolor[named]{Purple}{cmyk}{0.55,1,0,0.15}
\definecolor[named]{DarkBlue}{cmyk}{1,0.58,0,0.21}
\definecolor[named]{Red}{cmyk}{0,0.90,0.86,0}

\usepackage{pgfplots}
\usetikzlibrary{arrows}
\usetikzlibrary{arrows.meta}
\usetikzlibrary{patterns,patterns.meta}
\usepgfplotslibrary{groupplots}
\usepgfplotslibrary{statistics}

\newcommand\ans[1]{
\vspace{6pt}
\noindent
\doublebox{
    \begin{minipage}{0.96\linewidth}
      #1
    \end{minipage}
  }
}
\definecolor{mygreen}{RGB}{124, 205, 124}
\definecolor{mylightgreen}{RGB}{179, 238, 185}
\definecolor{myred}{RGB}{238, 162, 173}
\definecolor{mylightred}{RGB}{251, 233, 230}
\definecolor{myblue}{RGB}{92, 172, 238}
\definecolor{mydrawgray}{gray}{0.4}
\definecolor{myviolet}{RGB}{163, 50, 212}
\definecolor{myyellow}{RGB}{119, 94, 47}

\newcommand{\highlightred}[1]{\sethlcolor{mylightred}\hl{#1}}
\newcommand{\highlightgreen}[1]{\sethlcolor{mylightgreen}\hl{#1}}

\renewcommand{\paragraph}[1]{\vspace{0.4em}\noindent{\bf #1.}\hspace{4pt}}

\def \tool {\textsc{Aegis}\xspace}
\def \baseline {VRL\xspace}

\algsetup{linenosize=\small}
\SetKwInput{KwInput}{Require}
\SetAlFnt{\small}

\newcommand\realnumberstyle[1]{}
\makeatletter
\newcommand{\linecolor}[3]{
    {\realnumberstyle{#3}}
    \begingroup
    \lst@basicstyle
    \ifnum\value{lstnumber}=#1
        \color{#2}
    \else
        \color{white}
    \fi
    \rlap{\hspace*{\lst@numbersep}
    \color@block{\linewidth}{\ht\strutbox}{\dp\strutbox}
    }
    \endgroup
}
\makeatother

\makeatletter
\newcommand{\newlinecolor}[1]{
    \begingroup
    \lst@basicstyle
    \color{#1}
    \rlap{\hspace*{-\lst@numbersep}\hspace*{1ex}
    \color@block{\linewidth}{\ht\strutbox}{\dp\strutbox}
    }
    \endgroup
}
\makeatother

\let\origthelstnumber\thelstnumber
\makeatletter
\newcommand*\Suppressnumber{%
  \lst@AddToHook{OnNewLine}{%
    \let\thelstnumber\relax%
  }%
}

\newcommand\Reactivatenumber[1]{%
  \global\c@lstnumber#1%
  \global\advance\c@lstnumber\m@ne\relax%
  \lst@AddToHook{OnNewLine}{%
  \let\thelstnumber\origthelstnumber%
  }%
}
\makeatother

\AtBeginDocument{%
  \providecommand\BibTeX{{%
    \normalfont B\kern-0.5em{\scshape i\kern-0.25em b}\kern-0.8em\TeX}}}

\setcopyright{cc}
\setcctype{by}
\acmJournal{TOSEM}
\acmYear{2025} \acmVolume{1} \acmNumber{1} \acmArticle{1} \acmMonth{1} \acmPrice{}\acmDOI{10.1145/3773034}

\begin{document}

\title{Synthesizing Efficient and Permissive Programmatic Runtime Shields for Neural Policies}

\author{Jieke Shi}
\email{jiekeshi@smu.edu.sg}
\affiliation{%
  \institution{School of Computing and Information Systems, Singapore Management University}
  \country{Singapore}
}
\author{Junda He}
\email{jundahe@smu.edu.sg}
\affiliation{%
  \institution{School of Computing and Information Systems, Singapore Management University}
  \country{Singapore}
}
\author{Zhou Yang}
\email{zhou.yang@ualberta.ca}
\authornote{This work was done while the author was a research scientist at Singapore Management University. Zhou Yang is the corresponding author.}
\affiliation{%
  \institution{Department of Computing Science, University of Alberta \& Alberta Machine Intelligence Institute (Amii)}
  \country{Canada}
}
\author{{\DJ}or{\dj}e {\v{Z}}ikeli{\'c}}
\email{dzikelic@smu.edu.sg}
\affiliation{%
  \institution{School of Computing and Information Systems, Singapore Management University}
  \country{Singapore}
}
\author{David Lo}
\email{davidlo@smu.edu.sg}
\affiliation{%
  \institution{School of Computing and Information Systems, Singapore Management University}
  \country{Singapore}
}

\ccsdesc[500]{Software and its engineering~Software verification}
\ccsdesc[500]{Software and its engineering~Automatic programming}
\ccsdesc[500]{Computing methodologies~Artificial intelligence}





\begin{abstract}
  With the increasing use of neural policies in control systems, ensuring their safety and reliability has become a critical software engineering task. One prevalent approach to ensuring the safety of neural policies is to deploy programmatic runtime shields alongside them to correct their unsafe commands. However, the programmatic runtime shields synthesized by existing methods are either computationally expensive or insufficiently permissive, resulting in high overhead and unnecessary interventions on the system.
  To address these challenges, we propose \tool, a novel framework that synthesizes lightweight and permissive programmatic runtime shields for neural policies. \tool achieves this by formulating the seeking of a runtime shield as a sketch-based program synthesis problem and proposing a novel method that leverages counterexample-guided inductive synthesis and Bayesian optimization to solve it. To evaluate \tool and its synthesized shields, we use eight representative control systems and compare \tool with the current state-of-the-art. Our results show that the programmatic runtime shields synthesized by \tool can correct all unsafe commands from neural policies, ensuring that the systems do not violate any desired safety properties at all times. Compared to the current state-of-the-art, \tool's shields exhibit a 2.2$\times$ reduction in time overhead and a 3.9$\times$ reduction in memory usage, suggesting that they are much more lightweight. Moreover, \tool's shields incur an average of 1.5$\times$ fewer interventions than other shields, showing better permissiveness.
\end{abstract}

\keywords{Runtime Shielding, Deep Reinforcement Learning, Program Synthesis, Trustworthy Systems}

\maketitle

\input{sections/intro}
\input{sections/background}
\input{sections/methodology}

\input{sections/results}
\input{sections/discussion}
\input{sections/related}
\input{sections/conclusion}

\begin{acks}
  This research/project is supported by the National Research Foundation Singapore and DSO National Laboratories under the AI Singapore Programme (AISG Award No: AISG2-RP-2020-017). We would like to thank the anonymous reviewers for their valuable feedback.
\end{acks}

\balance
\bibliographystyle{ACM-Reference-Format}
\bibliography{reference}

\end{document}

%% file: sections/intro.tex
\section{Introduction}
\label{sec:intro}

Recent years have seen a notable increase in the use of deep neural networks (DNNs) to support control systems like autonomous vehicles~\cite{song2022cyber,ahmed2021machine,lopez2023arch}. In these systems, DNNs, also referred to as {\it neural policies}~\cite{zhu2019inductive,donti2021enforcing}, process information from the physical environment and send commands to hardware components like actuators to regulate their behavior in accordance with desired specifications. Neural policies excel at handling complex control tasks, such as autonomous driving~\cite{kiran2021deep, muhammad2020deep}, aerospace guidance~\cite{chai2021review, dong2021deep}, and even nuclear fusion~\cite{degrave2022magnetic,seo2024avoiding}, often outperforming traditional control software~\cite{song2022cyber, LillicrapHPHETS15,duan2016benchmarking,NIPS2017_9ddb9dd5}. According to Microsoft's projection, by 2025, 50\% of industrial companies will embrace control systems with neural policies, increasing productivity by 20\% and generating \$3.7 trillion in revenue~\cite{whitepaper}.

While neural policies have proven effective in various domains, concerns about their safety have arisen: they can occasionally produce unexpected or inappropriate commands, causing system failures that violate desired safety specifications~\cite{pang2022mdpfuzz,amodei2016concrete,raji2023concreteproblemsaisafety,he2024curiosity,pei2017deepxplore,song2022cyber,zhang2020machine}. This risk is exemplified in safety-critical tasks, e.g., autonomous driving, where neural policies may fail to control vehicles at desired heading angles, resulting in scenarios such as vehicles veering off the road or even colliding with pedestrians. Such catastrophic accidents have been reported multiple times in real-world systems employing neural policies, such as Tesla Autopilot~\cite{Tesla-crash,qzRodeTesla} and Waymo Driver~\cite{Waymos-crash,techcrunchWaymoSelfdriving}. In addition, certain neural policies in aerospace and manufacturing control systems have also demonstrated unsafe behavior~\cite{mashableAggressiveRiskier,telegraphCrushedDeath}, further highlighting the safety risks associated with using neural policies in safety-critical domains.

\sethlcolor{mylightgreen}

\begin{wrapfigure}{r}{0.56\linewidth}
    \centering
    \includegraphics[width=\linewidth]{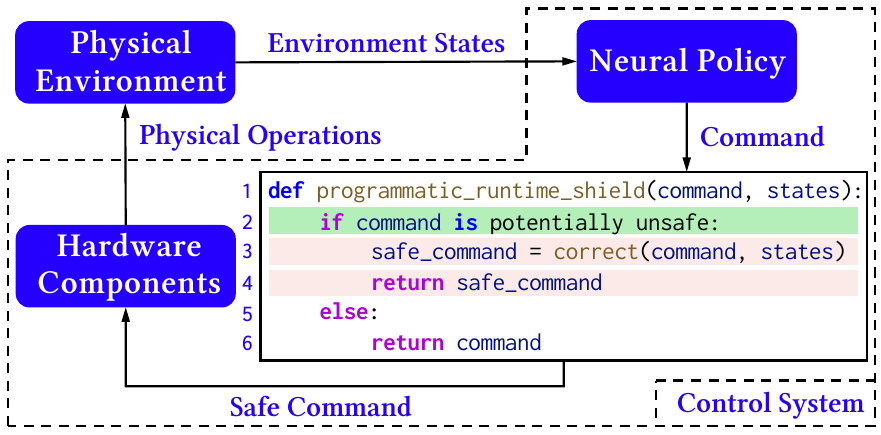}
    \caption{\small Overview of how a programmatic runtime shield works. The \hl{green} line detects potentially unsafe commands from the neural policy, while the {\setlength{\fboxsep}{1.1pt}\colorbox{mylightred}{red}} lines correct them by returning safe ones instead.}
    \label{fig:runtime-shielding}
\end{wrapfigure}

\paragraph{Programmatic Runtime Shields}
To date, programmatic runtime shields have demonstrated effectiveness in ensuring the safety of neural policies~\cite{zhu2019inductive,alshiekh2018safe,xiong2021scalable,guo2023rave}. These shields are typically simple and verifiable programs that can run alongside the neural policy and promptly correct any inappropriate commands from the policy as it executes. Figure~\ref{fig:runtime-shielding} illustrates how a programmatic runtime shield works within a control system using a neural policy. Specifically, the shield monitors commands from the neural policy and intervenes when it detects potentially unsafe commands that could cause the system to violate required safety properties (line 2, marked in \highlightgreen{green}). Upon intervention, the shield corrects unsafe commands by replacing them with safe alternatives based on the current state of the physical environment (lines~3-4, marked in \highlightred{red}). This ensures that the desired safety properties are maintained at all times.

Given that synthesizing these shields is generally less expensive than performing formal verification or exhaustive testing of neural policies, they offer a practical approach to ensuring their safety~\cite{zhu2019inductive, konighofer2020shield, xiong2021scalable}. Moreover, both the synthesis and operation of runtime shields typically rely only on the inputs and outputs of the policies, without requiring access to the neural network parameters. As a result, deploying runtime shields is often the only viable option for addressing safety violations of neural policies, particularly when retraining neural networks is prohibitively expensive or when accessing the parameters of the neural policies is restricted in the case of using proprietary products from third-party vendors.

\vspace{0.4em}
\noindent\textbf{What Makes Good Programmatic Runtime Shields?}\hspace{3pt}
Three critical attributes define the efficacy of programmatic runtime shields: {\it effectiveness}, {\it efficiency}, and {\it permissiveness}~\cite{zhu2019inductive,bloem2015shield,konighofer2017shield,wu2019shield,konighofer2020shield}. First, a good shield should effectively correct any unsafe command from the neural policy, thereby ensuring the perpetuation of desired safety properties in the system. Second, as an additional component running alongside the neural policy, the shield should be efficient, minimizing time and memory overhead. Lastly, maintaining permissiveness is essential~\cite{zhu2019inductive,konighofer2023online,konighofer2020shield}. The shield's judgment of whether a command from the neural policy is unsafe is not always accurate, and it may intervene unnecessarily when it encounters safe commands. The shield should be permissive, minimizing unnecessary intervention and allowing the neural policy to operate as freely as possible.

Deficiencies in any of these key attributes can have adverse consequences. For example, an inefficient runtime shield can consume excessive system resources, resulting in degraded system performance or crashes. This is particularly problematic in real-time systems~\cite{kopetz2022real} like autonomous vehicles, where timely responses are critical, or in resource-constrained embedded systems~\cite{roth2024resource,panda1999memory} like drones, where memory is limited. In addition, a non-permissive shield can often interfere with safe commands, thereby degrading system stability (i.e., a system's ability to achieve steady states in response to random inputs or disturbances, typically measured by required time steps~\cite{zhu2019inductive}; see Section~\ref{sec:discussion:runtime-shields-vs-neural-policies} for further discussion). Unfortunately, current state-of-the-art programmatic runtime shields still have limitations with respect to these key attributes, as we discuss next.

\paragraph{State-of-the-art and Its Limitations}
Several methods for synthesizing programmatic runtime shields have emerged in recent years, among which Zhu et al.'s work (An Inductive Synthesis Framework for Verifiable Reinforcement Learning~\cite{zhu2019inductive}, hereafter referred to as \baseline) stands out as the state-of-the-art. \baseline crafts programmatic runtime shields capable of effectively rectifying all unsafe commands to consistently uphold desired safety properties (see Section~\ref{sec:backgrounds} for more details). However, these shields exhibit limitations in the other two attributes. First, \baseline's shields involve operations with multiple high-degree polynomials, resulting in a considerable large time and memory overhead on the systems they operate within. Second, their lack of permissiveness often causes unnecessary interventions in otherwise safe commands, consequently compromising system stability. These limitations underscore opportunities for improving \baseline's shields, necessitating our work to advance the state-of-the-art.

\paragraph{Our Solution: \tool}
We introduce \tool, a novel method to synthesize efficient and permissive programmatic runtime shields for neural policies. The core idea of \tool is to formulate the seeking of a shield as a sketch-based program synthesis problem and use {\it counterexample-guided inductive synthesis} (CEGIS)~\cite{10.5555/1714168} and {\it Bayesian optimization} (BO)~\cite{garnett2023bayesian} to solve it. These two techniques have been studied in program synthesis~\cite{DBLP:conf/cav/AbateDKKP18,vcevska2019counterexample,saad2019bayesian,10.1007/978-3-031-65630-9_15} and related areas such as program repair~\cite{10.1145/3510418}, program translation~\cite{10.1145/3485489}, specification inference~\cite{10.1145/3720505,10.1145/3649844} and more~\cite{10.1145/3597503.3639190,10.1145/3632870}. However, \tool is the first to combine them to enhance the efficiency and permissiveness of runtime shields, offering a novel and unique perspective to the field.

Concretely, \tool begins with constructing a program sketch that implements a linear policy and a switching strategy. The linear policy is a linear function whose coefficients are parameters to be synthesized for ensuring the generation of safe commands. Meanwhile, the switching strategy formulates a linear inequality based on the disparity between the commands from the neural policy and the linear policy, along with a threshold (also a parameter to be synthesized), to determine when to invoke the linear policy for correcting unsafe commands. With the program sketch, \tool uses a novel counterexample-guided inductive synthesis algorithm and a Bayesian optimization algorithm to derive the necessary parameters for the linear policy and the switching strategy, respectively. Once all the parameters are obtained, \tool integrates them into the sketch, forming a programmatic runtime shield that runs alongside the neural policy to ensure safety.

Two thrusts make \tool superior to the state-of-the-art. First, the shields synthesized by \tool involve only linear functions or linear inequalities instead of the high-degree polynomials in \baseline's shields. This design choice reduces computational complexity and overhead, making our shields much more efficient. Second, our Bayesian optimization algorithm assigns the optimal threshold to the switching strategy by minimizing unnecessary interventions, thus making our shields more permissive than those synthesized by \baseline.

\paragraph{Evaluating \tool}
We evaluate \tool primarily on the three key attributes of programmatic runtime shields: effectiveness, efficiency, and permissiveness. In addition, we also evaluate \tool's scalability, showing its ability to efficiently synthesize shields for neural policies of different sizes. Our evaluation uses 4 representative control systems and compares \tool to \baseline.

Our results show that the programmatic runtime shields synthesized by \tool effectively correct all unsafe commands from neural policies, ensuring that the systems do not violate the desired safety properties at any time. Notably, compared to \baseline, \tool's shields exhibit a 2.2$\times$ reduction in time overhead and a 3.9$\times$ reduction in memory overhead, suggesting they are much more efficient. \tool's shields also show improved permissiveness, incurring an average of 1.5$\times$ fewer interventions than those synthesized by \baseline. We also gauge the ratio of necessary interventions to all interventions and find that, on average, our shields intervene when necessary 10.2\% more frequently than \baseline's. Also, \tool shows superior scalability relative to \baseline, reducing synthesis time by 3.0$\times$. Additionally, we provide an analysis of replacing neural policies with \tool's shields, confirming the necessity for the coexistence of neural policies and shields to maintain the system's safe and stable operation.

\paragraph{Novelty and Contributions}
While \tool leverages two established methodological paradigms, i.e., counterexample-guided inductive synthesis (CEGIS) and Bayesian optimization (BO), its novelty lies in how these paradigms are instantiated, customized, and tightly integrated for the specific task of synthesizing runtime shields for neural policies. Unlike prior shield synthesis works~\cite{zhu2019inductive,10.1007/978-3-030-81685-8_22} or even other applications of CEGIS in broader domains~\cite{10.1145/3510418,sketch08,solar2009sketching,9978938,edwards2024fossil,10.1007/978-3-031-65630-9_15,10.1145/3632870,10.1145/3485489}, \tool uniquely formulates the synthesis problem as a sketch-based search over linear templates and introduces a novel refinement algorithm that leverages the maximal output admissible set to identify counterexamples and refine the shields (see Section~\ref{sec:counterexample-guided-synthesis} for more details). Furthermore, we are the first to incorporate BO into shield synthesis, tailoring it with a domain-specific objective function to optimize shields' permissiveness by selecting appropriate switching thresholds (see Section~\ref{sec:switching-strategy-synthesis} for more details). In addition, to the best of our knowledge, \tool is the first to combine CEGIS and BO into a unified framework for shield synthesis, which we believe is both technically novel and offers a unique perspective on the safety assurance of neural controllers.

\vspace{0.4em}

To summarize, our main contributions are as follows:

\begin{itemize}[topsep=0.4em]
\item We improve both the efficiency and permissiveness of programmatic runtime shields for neural policies, addressing key limitations of the state-of-the-art.
\item We propose and implement \tool, a novel synthesis method that formulates runtime shield generation as sketch-based program synthesis, and is the first to combine CEGIS and BO for this purpose.
\item We evaluate \tool on eight representative control systems and show its superiority over the state-of-the-art, with significant improvements in all metrics.
\end{itemize}

\paragraph{Paper Structure}
The rest of the paper is organized as follows. Section~\ref{sec:backgrounds} covers the preliminary information of our work, and Section~\ref{sec:approach} describes in detail how \tool works. Section~\ref{sec:results} presents the experimental setup and analyzes the experimental results. Section~\ref{sec:discussion} provides additional discussions about our work, as well as threats to validity. Section~\ref{sec:related} presents related work. Section~\ref{sec:conclusion} concludes our paper, and presents our future work and replication package.

%% file: sections/background.tex
\section{Preliminary}
\label{sec:backgrounds}

This section provides a brief introduction to neural policies, as well as the synthesis of programmatic runtime shields.

\paragraph{Neural Policy}
Recent advances in neural networks have spurred their adoption in many control systems~\cite{song2022cyber,ahmed2021machine,lopez2022arch}. These neural networks generate commands to control systems that interact with the physical environment to fulfill desired specifications, often referred to as neural policies~\cite{zhu2019inductive}. For clarity, below we formally describe neural policies.

\sethlcolor{mylightgreen}
\begin{wrapfigure}{r}{0.48\linewidth}
    \centering
    \includegraphics[width=\linewidth]{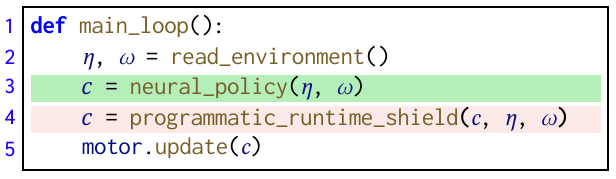}
    \caption{\small Main control loop in a quadcopter (simplified for readability). The \hl{green} and \highlightred{red} colors highlight the neural policy and runtime shield, respectively.}
    \vspace{-1.5em}
    \label{fig:example-main-loop}
  \end{wrapfigure}

We begin by defining the physical environment as $\mathcal{E} = \langle \mathcal{S}, \mathcal{S}_0, \mathcal{S}_{u}, \mathcal{C}, f, r\rangle $. Here $\mathcal{S}$ is the set of all possible states of the environment, with $s \in \mathcal{S}$ denoting a single state. The subset $\mathcal{S}_0 \subset \mathcal{S}$ specifies the initial states from which systems working in this environment can commence. $\mathcal{S}_u \subset \mathcal{S}$ is a set of unsafe states that systems must avoid, thereby defining the desired safety specifications. $\mathcal{C}$ represents a set containing all feasible control commands that neural policies can issue, where $c \in \mathcal{C}$ denotes an individual command. The environment dynamics function $f$ governs the transition from one environment state to another upon receiving a command from a policy. Additionally, the reward function $r$ maps a state and a command to a real value as a feedback signal to guide the training process of a neural policy.

Modern control systems are typically implemented with digital rather than analog techniques~\cite{lee2015past,zhu2019inductive}, where signals are processed discretely over fixed time intervals. Thus, the state transition of an environment can be formulated with discrete and fixed time step $\Delta t$ using Euler's method~\cite{hildebrand1987introduction,zhu2019inductive}. Namely, a state $s_{t}$ transitions to the subsequent state $s_{t+1}$ after time $t$ by the following equation:
\begin{equation}
\label{eq:transition}
    s_{t+1} = s_{t} + f(s_{t}, c_t) \times \Delta t
\end{equation}
Here, $c_t$ denotes a command generated by a neural policy $\pi$, which is implemented as a neural network that takes an environment state $s_t$ as input, i.e., $c_t = \pi(s_t)$, and $\Delta t$ is the fixed time step.
Typically, neural policies are trained using Deep Reinforcement Learning (DRL) algorithms~\cite{10.1007/978-3-319-56991-8_32, LillicrapHPHETS15}, which optimize the policy's parameters to maximize the expected discounted cumulative reward through interaction with the environment. Specifically, in an environment $\mathcal{E}$, a system begins at an initial state $s_0 \in \mathcal{S}_0$, and its neural policy produces commands to control the system to interact with the environment until it reaches a termination state, $s_T$, after $T$ time steps. The interaction history is called a trajectory:
\begin{equation}
    \tau: (\langle s_0,c_0 \rangle, \cdots, \langle s_t,c_t \rangle, \langle s_{t+1},c_{t+1} \rangle \cdots, \langle s_{T},c_{T} \rangle)
\end{equation}
The discounted cumulative reward is the sum of the rewards multiplied by a discount factor $\gamma \in [0,1]$ over a trajectory, denoted as $R(\tau) = \sum_{i=0}^T r(s_i,c_i) \cdot \gamma^i$. The training objective of DRL is to find the optimal neural policy $\pi^\ast$ that maximizes the expected discounted cumulative reward over all trajectories, which can be formalized as follows:
\begin{equation}
    \pi^\ast = \arg\max_{\pi} \mathbb{E}_{\tau \sim \pi} \left[ R(\tau) \right]
\end{equation}
The specific DRL algorithms for solving the above equation vary with different tasks and are beyond our paper's scope. We refer interested readers to~\cite{10.1007/978-3-319-56991-8_32,LillicrapHPHETS15} for more details.

After training, the neural policy $\pi$ works to control the system on which it was trained. Figure~\ref{fig:example-main-loop} shows an example of the main control loop of a quadcopter using a neural policy. The control loop starts by retrieving the current state, i.e., the current attitude angle $\eta$ and angular velocity~$\omega$ (line~2). Next, the state is fed into the neural policy to obtain the resulting commands (line~3, marked in \highlightgreen{green}). The command is then transmitted to the quadcopter's motor to adjust its attitude and angular velocity (line~5). This loop continuously iterates to drive the quadcopter.

\paragraph{Programmatic Runtime Shields and \baseline}
While neural policies outperform traditional control methods in adapting control systems to more complex tasks~\cite{LillicrapHPHETS15,hunt1992neural,duan2016benchmarking}, there remain potential risks that neural policies may output unsafe commands, resulting in system failures, such as misregulating motor power to cause quadcopters to crash. Therefore, a runtime shield becomes imperative to collaborate with the neural policy, ensuring the system's safety when the neural policy fails to do so. Line~4 (marked in \highlightred{red}) in Figure~\ref{fig:example-main-loop} shows the integration of a programmatic runtime shield into a control loop. The shield processes the neural policy's command and the current state as input, generating a safe command as a replacement if the neural policy's command is deemed unsafe.

\begin{figure}[t]
    \centering
    \begin{minipage}[c]{0.5\linewidth}
      \centering
      \includegraphics[width=\linewidth]{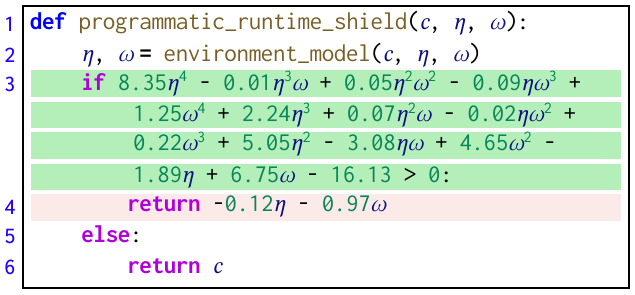}
      \captionof{figure}{\small A runtime shield synthesized by \baseline for the neural policy in a quadcopter system. Line~3 (marked in \highlightgreen{green}) is a barrier certificate that identifies unsafe states, while a linear policy in line~4 (marked in \highlightred{red}) outputs safe commands.}
    \label{fig:example-baseline}
  \end{minipage}
  \hfill
  \begin{minipage}[c]{0.46\linewidth}
    \centering
    \includegraphics[width=\linewidth]{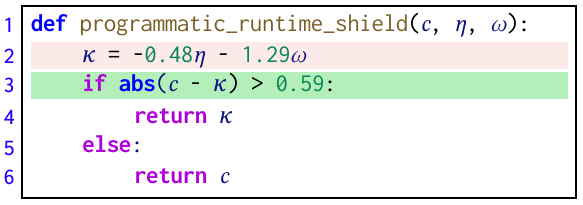}
    \captionof{figure}{\small A runtime shield synthesized by \tool for the neural policy in a quadcopter system. Line~2 (marked in \highlightred{red}) is a linear policy that outputs safe commands, while line~3 (marked in \highlightgreen{green}) identifies potentially unsafe commands.}
    \label{fig:example-tool}
  \end{minipage}
\end{figure}

\baseline~\cite{zhu2019inductive} is a state-of-the-art approach for synthesizing programmatic runtime shields. Figure~\ref{fig:example-baseline} shows an example of its shield for a neural policy within a quadcopter system. The shield contains a high-degree polynomial expression in line~3 (marked in \highlightgreen{green}), which is a {\it barrier certificate}~\cite{prajna2004safety,gulwani2008constraint,kong2013exponential}, along with a linear policy in line~4 (marked in \highlightred{red}) that has been verified to generate safe commands. The barrier certificate is a proof that if its result for a given state is non-positive, then the system is safe in that state; otherwise, the system is unsafe. When \baseline's shield works, it first uses a known mathematical model of the environment to predict the next state with the neural policy's command~(line~2), and then checks the safety of the next state using the barrier certificate (line~3). If the certificate yields a positive result, entering the next state steered by the neural policy's command can inevitably make the system unsafe. In such cases, the shield returns the command of the linear policy that has a guarantee of safety instead (line~4). Conversely, if the certificate's result is non-positive, the shield has the neural policy's command returned (line~6).

\sethlcolor{mylightgreen}

As discussed in Section~\ref{sec:intro}, the runtime shields created by \baseline still have limitations regarding efficiency and permissiveness, as evidenced by their high runtime overhead and frequent unnecessary interventions in safe commands. Therefore, we introduce \tool to improve the state-of-the-art. To give a sense of how our shields work compared to \baseline's, Figure~\ref{fig:example-tool} shows a shield synthesized by \tool. The conditional statement in line~3 (marked in \highlightgreen{green}) is a linear inequality, which offers improved computational efficiency over \baseline's barrier certificate with high-degree polynomials. On the other hand, the linear inequality measures how far away the neural policy output is from the output of the safe linear policy (line~2, marked in \highlightred{red}), which has been formally verified. This direct comparison of the neural policy output to a safe output, rather than to an output of a polynomial barrier function that was fitted to bound the safe outputs, leads to an improved permissiveness of shields computed by our approach.

%% file: sections/methodology.tex
\section{\tool}
\label{sec:approach}

\subsection{Overview}

\begin{figure*}[t!]
    \centering
    \includegraphics[width=\linewidth]{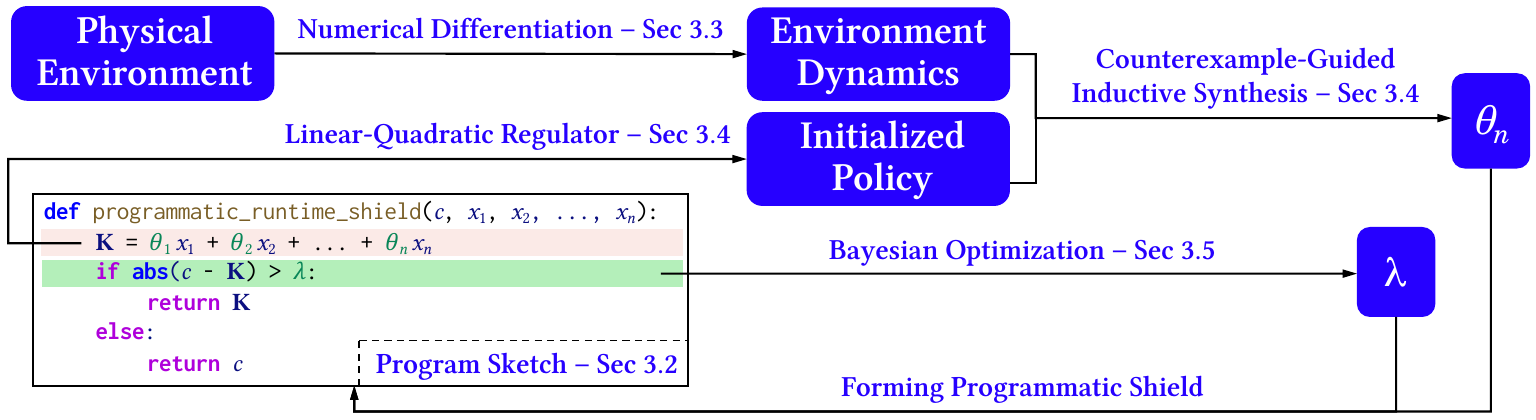}
    \caption{The workflow of \tool.}
    \label{fig:overview}
\end{figure*}

\begin{wrapfigure}{r}{0.465\linewidth}
    \centering
    \includegraphics[width=0.9\linewidth]{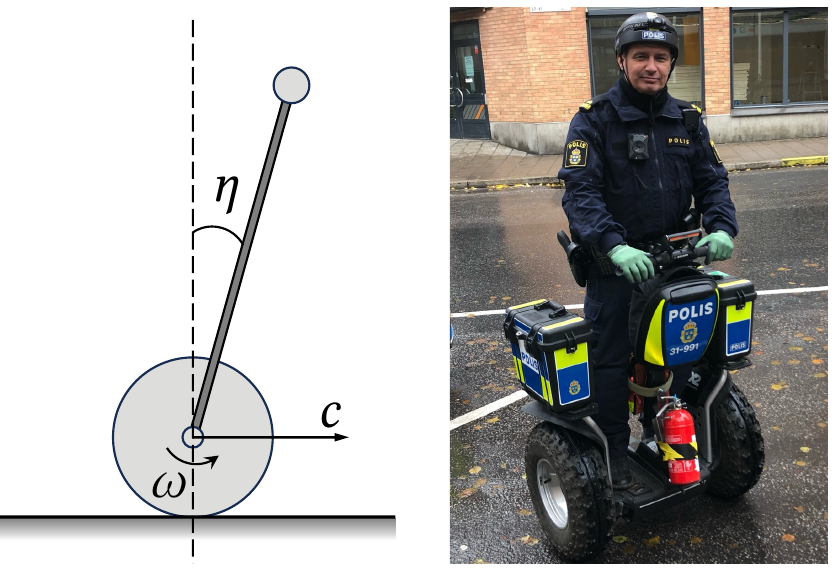}
    \caption{Pendulum system (left) used as a model for the Segway personal transporter (right); image from Wikipedia~\cite{segwayPoliceSweden2020}. $\eta$ is the pendulum's angle, $\omega$ is the angular velocity, and $c$ is the command from a neural policy to adjust wheel motion.}
    \label{fig:example-cartpole}
    \vspace{-0.5em}
  \end{wrapfigure}

This section provides details of \tool, with its workflow illustrated in Figure~\ref{fig:overview}. \tool begins by constructing a program sketch that consists of a linear policy and a switching strategy (Section~\ref{sec:program-sketch}). Then, given a complex and possibly unknown physical environment, \tool derives its state transition function by numerical differentiation~\cite{griewank2013stable} (Section~\ref{sec:state-transition}). Next, \tool initializes a linear policy using the Linear-Quadratic Regulator (LQR) algorithm~\cite{anderson2007optimal} and iteratively refines it with a counterexample-guided inductive synthesis algorithm, until the linear policy is verified to be safe against the safety specifications (Section~\ref{sec:counterexample-guided-synthesis}), determining the parameters of the linear policy. Following this, \tool synthesizes the threshold of the switching strategy using a Bayesian optimization algorithm, aiming to minimize unnecessary shield interventions (Section~\ref{sec:switching-strategy-synthesis}). Finally, the synthesized components are incorporated into the program sketch to form a programmatic runtime shield. The following subsections describe each step in detail.

\paragraph{An Illustrative Example}
To concretely demonstrate \tool's synthesis process, we adopt the pendulum system as a running example. This classic benchmark abstracts a rigid pendulum pivoted at the base and serves as a simplified model for many real-world control systems that require high assurance but remain difficult to verify~\cite{zhu2019inductive,guo2023rave,xiong2021scalable}. In our example, we use it to model the widely-used Segway personal transporter, as shown in Figure~\ref{fig:example-cartpole}. Here, the rider and the steering column are represented as a pendulum mounted on wheels. The system's state consists of the angle $\eta$, representing the deviation of the pendulum from the vertical axis, and the angular velocity $\omega$, denoting the changing rate of this angle. A neural policy produces control commands $c$ to maintain balance by controlling the motion of the wheels, which in turn affects the pendulum's orientation. Following Zhu et al.~\cite{zhu2019inductive}, the system starts from a set of initial states $\mathcal{S}_0$:
\begin{equation}
    \mathcal{S}_0 = \left\{s \in \mathcal{S} \mid |\eta| \leq 20^{\circ} \wedge |\omega| \leq 20^{\circ}/\text{s} \right\}
\end{equation}
The safety property we wish to preserve is that the pendulum never falls into an unrecoverable state nor swings excessively, which can be quantitatively specified as constraining the angle $\eta$ to remain within $30^\circ$ and the angular velocity $\omega$ within $30^\circ/\text{s}$. Formally, these constraints, i.e., the set of safe states $\neg \mathcal{S}_u$, can be expressed as:
\begin{equation}
    \neg \mathcal{S}_u = \left\{s \in \mathcal{S} \mid |\eta| < 30^{\circ} \wedge |\omega| < 30^{\circ}/\text{s} \right\}
\end{equation}

\begin{wrapfigure}{r}{0.5\linewidth}
    \centering
    \includegraphics[width=\linewidth]{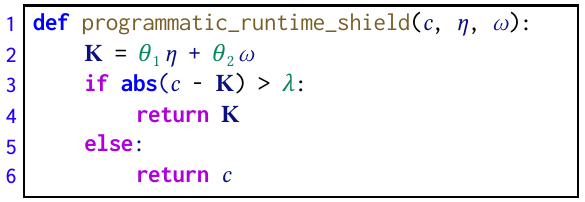}
    \caption{\tool's program sketch for the pendulum system. $\theta_1$, $\theta_2$, and $\lambda$ are the parameters to be synthesized.}
    \vspace{-0.5em}
    \label{fig:sketch-example}
\end{wrapfigure}

As illustrated in Figure~\ref{fig:overview}, the first step of \tool involves constructing a program sketch for the runtime shield. In this example, the sketch is depicted in Figure~\ref{fig:sketch-example}. It takes as input the control command $c$ produced by the neural policy and the environment state $(\eta, \omega)$, and returns a command that satisfies safety constraints. Line~2 defines a linear policy that generates a safe command $\mathbf{K}$, where the coefficients $\theta_1$ and $\theta_2$ are to be synthesized. Line~3 specifies a switching strategy that determines when to replace the potentially unsafe command $c$ with the linear policy's command $\mathbf{K}$, with the switching threshold $\lambda$ also to be synthesized.

\tool then infers the unknown environment dynamics function $f(\eta, \omega, c)$, which describes how the pendulum system responds to commands from the neural policy in a given state. It approximates the dynamics as a linear time-invariant system and derives a linear approximation $f'(\eta, \omega, c)$ using numerical differentiation~\cite{griewank2013stable}, as shown below (more details in Section~\ref{sec:state-transition}):
\begin{equation}
    f'(\eta_t, \omega_t, c_t) =
    \begin{bmatrix}
    1.9027 & -1 \\
    1 & 0
    \end{bmatrix}
    \begin{bmatrix}
    \eta_t \\
    \omega_t
    \end{bmatrix}
    +
    \begin{bmatrix}
    1 \\
    0
    \end{bmatrix}
    c_t
    \approx f(\eta_t, \omega_t, c_t)
    \end{equation}
Thus, the state transition function, which describes how the pendulum system evolves from the current state $(\eta_t, \omega_t)$ at time $t$ to the subsequent state $(\eta_{t+1}, \omega_{t+1})$ under the command $c_t$ from the neural policy, can be written as:
\begin{equation}
    \begin{bmatrix}
    \eta_{t+1} \\
    \omega_{t+1}
    \end{bmatrix}
    \approx
    \begin{bmatrix}
    \eta_t \\
    \omega_t
    \end{bmatrix}
    +
    f'(\eta_t, \omega_t, c_t)\, \Delta t
    =
    \begin{bmatrix}
    \eta_t \\
    \omega_t
    \end{bmatrix}
    +
    \left(
    \begin{bmatrix}
    1.9027 & -1 \\
    1 & 0
    \end{bmatrix}
    \begin{bmatrix}
    \eta_t \\
    \omega_t
    \end{bmatrix}
    +
    \begin{bmatrix}
    1 \\
    0
    \end{bmatrix}
    c_t
    \right) \Delta t
    \end{equation}
where $\Delta t$ is the fixed time step.

\begin{wrapfigure}{r}{0.5\linewidth}
    \centering
    \includegraphics[width=\linewidth]{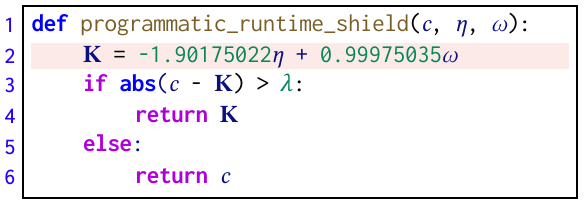}
    \caption{The shield with the initialized linear policy.}
    \label{fig:linear-policy}
    \vspace{-0.5em}
\end{wrapfigure}

With the inferred dynamics, \tool initializes the linear policy via the Linear-Quadratic Regulator (LQR) algorithm~\cite{anderson2007optimal}, which computes the policy parameters $\theta_1 = -1.90175022$ and $\theta_2 = 0.99975035$, as shown in line~2 (highlighted in \highlightred{red}) of Figure~\ref{fig:linear-policy}. Further computational details are provided in Section~\ref{sec:counterexample-guided-synthesis}.

\begin{figure}[h]
    \centering
    \begin{minipage}[c]{0.49\linewidth}
      \centering
      \includegraphics[width=\linewidth]{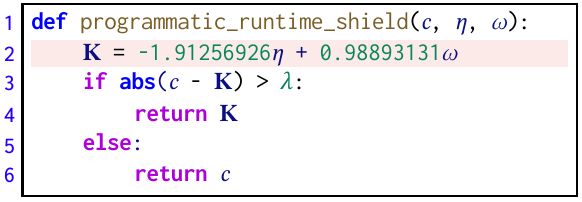}
      \captionof{figure}{\small The shield with the refined linear policy.}
      \label{fig:refined-linear-policy}
  \end{minipage}
  \hfill
  \begin{minipage}[c]{0.49\linewidth}
    \centering
    \includegraphics[width=\linewidth]{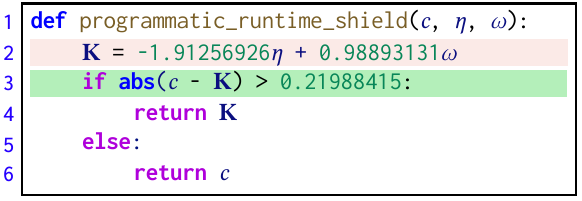}
    \captionof{figure}{\small The final runtime shield synthesized by \tool.}
    \label{fig:runtime-shield}
  \end{minipage}
\end{figure}

Then, \tool verifies the linear policy's safety by computing the Maximal Output Admissible Set (MOAS)~\cite{gilbert1991linear} (see Section~\ref{sec:counterexample-guided-synthesis} for details). The MOAS identifies the initial states that can be safely controlled by the linear policy, i.e., states in $\mathcal{S}_0$ from which the system can avoid entering unsafe states $\mathcal{S}_u$. If the MOAS does not fully encompass $\mathcal{S}_0$, this indicates that the linear policy is unsafe, as some initial states may lead the system into unsafe regions. In such cases, the uncovered states are treated as counterexamples to guide policy refinement.
For the pendulum example, in the first iteration, the boundary state with $\eta = 20^{\circ}$ and $\omega = 20^{\circ}/\text{s}$ lies outside the MOAS (the full MOAS expression is omitted here for brevity), demonstrating that the initial policy is insufficiently safe. \tool then refines the policy using the algorithm described in Section~\ref{sec:counterexample-guided-synthesis}, adjusting $\theta_1$ and $\theta_2$ to expand the MOAS until it fully covers $\mathcal{S}_0$. After several refinements, the final linear policy is obtained with $\theta_1 = -1.91256926$ and $\theta_2 = 0.98893131$, as shown in line 2 (highlighted in \highlightred{red}) of Figure~\ref{fig:refined-linear-policy}.

The next step is to synthesize the switching strategy, which introduces a threshold $\lambda$ that determines when to replace the command $c$ from the neural policy with the command $\mathcal{K}$ from the linear policy. If this threshold is not properly set, the system may remain unsafe, as the verified safe linear policy would not be invoked to override unsafe commands. \tool synthesizes this threshold using a Bayesian optimization algorithm (see Section~\ref{sec:switching-strategy-synthesis} for details), which aims to minimize safety violations while also limiting unnecessary interventions by the linear policy, thereby preserving the system's stability. The synthesized threshold is $0.21988415$, meaning that if the commands from the neural and linear policies differ by more than $0.21988415$, the system will use the linear policy's command instead. The final runtime shield is shown in Figure~\ref{fig:runtime-shield}, where line~3 (highlighted in \highlightgreen{green}) specifies the synthesized switching strategy.

\subsection{Program Sketch}
\label{sec:program-sketch}

\begin{wrapfigure}{r}{0.5\linewidth}
    \centering
    \includegraphics[width=\linewidth]{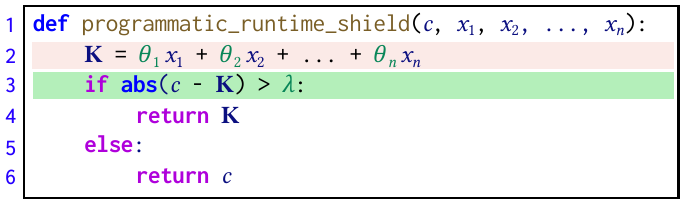}
    \caption{The universal program sketch of a runtime shield synthesized by \tool.}
    \label{fig:sketch}
  \end{wrapfigure}

In \tool, we model the synthesis of programmatic runtime shields as a sketch-based program synthesis problem~\cite{solar2009sketching, 10.5555/1714168}. A sketch is a partial code template with blank holes to be filled by synthesized parameters or statements. Considering that the synthesized shields should impose minimal overhead on the system, we set our program sketch to include only linear functions and inequalities, rather than following Zhu et al.'s work~\cite{zhu2019inductive} to use barrier certificates with high-order polynomials. This is because linear functions and inequalities are more efficient in terms of computational cost and are also widely used in control systems~\cite{mania2018simple}. Another advantage of using linear functions is that the computational cost of synthesizing them is relatively lower compared to obtaining barrier certificates by solving optimization problems with constraint solvers~\cite{zhu2019inductive,10.1145/3106237.3106281}.

We construct the program sketch as Figure~\ref{fig:sketch}. This sketch partially implements a runtime shield that takes the command $c$ from the neural policy and the environment state $x_1, x_2, ..., x_n$ as input and returns a safe command. There is a linear policy that produces a command $\mathbf{K}$ based on the environment state variables (line~2, marked in \highlightred{red}), where $\theta_1, \dots, \theta_n$ are coefficients in the linear policy to be synthesized. Another main component is a conditional statement, which is a switching strategy that determines when to invoke the linear policy to replace the unsafe command $c$ with the safe command $\mathbf{K}$ (line~3, marked in \highlightgreen{green}), where $\lambda$ is the threshold in the switching strategy to be synthesized. The sketch is then instantiated by filling the holes with the parameters synthesized by \tool.

Here, we formulate the shield using a linear policy not only to align with prior work~\cite{zhu2019inductive}, but also due to its ability to approximate key behaviors of neural policies, efficiency, and compatibility with formal safety analysis. Prior studies have demonstrated that linear policies can capture the core functionalities of more complex controllers~\cite{mania2018simple,zhu2019inductive,10.1145/3715105}. For instance, Rajeswaran et al.~\cite{NIPS2017_9ddb9dd5} show that while robots controlled by linear policies may not walk like those using neural policies, they can still stand effectively, achieving critical safety-relevant behaviors. Furthermore, linear policies execute significantly faster~\cite{10.1145/3715105}, which is crucial in runtime shielding, where minimizing overhead is essential to maintaining system responsiveness—especially in safety-critical applications where even minor delays in executing commands can have serious consequences. In addition, linear policies enable the use of easily tractable verification techniques, such as maximal output admissible sets~\cite{gilbert1991linear} or barrier certificates~\cite{prajna2004safety}, making them well-suited for integration into our synthesis framework. While exploring more complex controllers is an interesting direction, doing so would necessitate a complete redesign of the synthesis pipeline, which we leave to future work.

\subsection{Inferring Environment Dynamics}
\label{sec:state-transition}

In real-world applications, the environments in which neural policies operate are often complex and characterized by unknown dynamics functions~\cite{10.5555/2968826.2968946}, making it challenging or infeasible to directly verify the effectiveness of the synthesized programmatic runtime shields~\cite{zhu2019inductive}. Therefore, \tool infers the dynamics function of the physical environment in which the neural policy operates, making the unknown environment dynamics explicit and thereby facilitating the subsequent synthesis and verification of the linear policy.

As defined in Section~\ref{sec:backgrounds}, an environment state $s_t$ transitions to the subsequent state $s_{t+1}$ based on the state transition function $s_{t+1} = s_t + f(s_t, c_t) \times \Delta t$, where $c_t$ represents the commands issued by the neural policy at time step $t$. What we aim to infer is $f(s_t, c_t)$, which describes the dynamics of the physical environment, i.e., how the next state evolves in response to both the commands from the neural policy and the current state. We adopt the assumption that the environment dynamics can be approximated by a discrete-time linear time-invariant system model~\cite{rugh1996linear,zadeh2008linear}, which is widely used for modeling the dynamics of physical environments~\cite{menghi2020approximation,zhu2019inductive,LillicrapHPHETS15}. Such a model can be defined as follows:
\begin{equation}
    \label{eq:linear-dynamics}
    f'(s_t, c_t) = \mathbf{A} s_t + \mathbf{B} c_t \approx f(s_t, c_t)
\end{equation}
where $\mathbf{A}$ and $\mathbf{B}$ are two constant matrices. The linear time-invariant system model $f'(s_t, c_t)$ is a linear approximation of the unknown dynamics function $f(s_t, c_t)$ around an equilibrium point~\cite{cooke1996problem,ramos2005linearization}, defined by specific environment states and commands $s^*$ and $c^*$ such that applying $c^*$ to $s^*$ results in no change in the environment state over time, i.e., $f(s^*, c^*) = 0$. In linearization, the equilibrium point is often chosen as the initial environment state $s_0$ and the command $c_0$ at the beginning of the system operation, i.e., $s^* = s_0$ and $c^* = c_0$, so that the system's behavior can be approximated by a linear model right from the start, simplifying the subsequent synthesis and verification processes.
In such a case, the matrices $\mathbf{A}$ and $\mathbf{B}$ in $f'(s_t, c_t)$ can be derived as the Jacobian matrices~\cite{arrowsmith1992dynamical}, which represent the first-order partial derivatives of $f(s_t, c_t)$ with respect to $s_t$ and $c_t$, respectively. Considering $f(s_t, c_t)$ as a vector-valued function with $s_t$ and $c_t$ as input vectors having $m$ and $n$ dimensions and $f(s_t, c_t)$ as a $z$-dimensional output vector, respectively, then the Jacobian matrices $\mathbf{A}$ and $\mathbf{B}$ are $z \times m$ and $z \times n$ matrices, respectively, with the elements defined as:
\begin{equation}
    \begin{gathered}
    \mathbf{A}_{i j}=\frac{\partial f_i\left(s_t, c_t\right)}{\partial s_j}, \quad \mathbf{B}_{i k}=\frac{\partial f_i\left(s_t, c_t\right)}{\partial c_k} \\
    i=1, \ldots, z, \quad j=1, \ldots, m, \quad k=1, \ldots, n
    \end{gathered}
\end{equation}
where $f_i(s_t, c_t)$ denotes the $i$-th element of the output vector $f(s_t, c_t)$, $s_j$ and $c_k$ are the $j$-th element of the state vector $s_t$ and the $k$-th element of the command vector $c_t$, respectively. The $\mathbf{A}_{i j}$ and $\mathbf{B}_{i k}$ are the elements of the Jacobian matrices $\mathbf{A}$ and $\mathbf{B}$, respectively.$m, l, u, M, x, \theta$

We obtain $\mathbf{A}$ and $\mathbf{B}$ using \emph{numerical differentiation}~\cite{gautschi2011numerical,cullum1971numerical}, which is commonly used to approximate a function's derivative from a limited set of input-output pairs~\cite{cheng2007numerical}. Specifically, it estimates the derivative by evaluating the function at two distinct inputs, computing the difference between the two outputs, dividing the difference by the distance between the two inputs, and then repeating the process for each input dimension. Formally, given the unknown dynamics function $f(s_t, c_t)$, the equilibrium point $s_0$ and $c_0$, the linearization algorithm computes the matrices $\mathbf{A}$ and $\mathbf{B}$ as follows:
\begin{equation}
    \label{eq:numerical-differentiation}
    \begin{gathered}
        \mathbf{A}_{i j} \approx \frac{f_i\left(s_0 + \epsilon e_j, c_0\right) - f_i\left(s_0 - \epsilon e_j, c_0\right)}{2 \times \epsilon}, \\
        \mathbf{B}_{i k} \approx \frac{f_i\left(s_0, c_0 + \epsilon e_k\right) - f_i\left(s_0, c_0 - \epsilon e_k\right)}{2 \times \epsilon} \\
    i=1, \ldots, z, \quad j=1, \ldots, m, \quad k=1, \ldots, n
    \end{gathered}
\end{equation}
where $e_j$ and $e_k$ are unit vectors with the $j$-th and $k$-th elements being 1 and the rest being 0, respectively, and $\epsilon$ is a small random perturbation.

\begin{algorithm}[t!]
     \small
    \SetKwInput{Input}{Require}
    \SetKwComment{Comment}{\textbf{**//**} }{}
    \SetCommentSty{upshape}
    \Input{$f$, $s_0$, $c_0$, $z$, $m$, $n$}
    $\mathbf{A} \gets \mathbf{0}$, $\mathbf{B} \gets \mathbf{0}$, $\epsilon \gets \text{random\_perturbation}()$\label{line:linearization-2}\\
    \For{$i \gets 1$ \KwTo $z$}{\label{line:linearization-3}
    \For{$j \gets 1$ \KwTo $m$}{
    $\mathbf{A}_{i j} \gets \big(f_i(s_0 + \epsilon e_j, c_0) - f_i(s_0 - \epsilon e_j, c_0)\big)/(2 \times \epsilon)$
     }
    \For{$k \gets 1$ \KwTo $n$}{
    $\mathbf{B}_{i k} \gets \big(f_i(s_0, c_0 + \epsilon e_k) - f_i(s_0, c_0 - \epsilon e_k)\big)/(2 \times \epsilon)$
     }
     }\label{line:linearization-4}
    \Return $\mathbf{A}, \mathbf{B}$ \label{line:linearization-5}
    \caption{Inferring Environment Dynamics}
    \label{alg:linearization}
    \end{algorithm}

Algorithm~\ref{alg:linearization} shows the pseudocode of how \tool infers the linear approximation $f'(s_t, c_t)$ of the unknown environment dynamics function via numerical differentiation. The algorithm takes as input the unknown dynamics function $f(s_t, c_t)$ (only its inputs and outputs can be accessed), the environment state $s_0$ and the command $c_0$ at the equilibrium point, the number of output dimensions $z$, the number of state dimensions $m$, and the number of command dimensions $n$. The algorithm first randomly produces a small perturbation $\epsilon$ (line~\ref{line:linearization-2}) to slightly alter the environment state $s_0$ and the command $c_0$ around the equilibrium point. Subsequently, Equation~\ref{eq:numerical-differentiation} is applied to compute the elements of the Jacobian matrices $\mathbf{A}$ and $\mathbf{B}$ by evaluating the unknown dynamics function $f(s_t, c_t)$ at the perturbed inputs (lines~\ref{line:linearization-3}-\ref{line:linearization-4}). Finally, the algorithm returns the matrices $\mathbf{A}$ and $\mathbf{B}$ (line~\ref{line:linearization-5}) to form the linear approximation $f'(s_t, c_t)$ of the dynamics function $f(s_t, c_t)$.

\subsection{Counterexample-Guided Inductive Synthesis}
\label{sec:counterexample-guided-synthesis}

With the inferred environment dynamics function, we are now prepared to synthesize the linear policy, which is the first major component of the programmatic runtime shield we create. Our goal is to synthesize a linear policy that ensures all outputs adhere to the safety specifications. To achieve this, we introduce a counterexample-guided inductive synthesis algorithm, which is a popular approach for synthesizing programs from specifications~\cite{vcevska2019counterexample, solar2009sketching, 10.5555/1714168}. This algorithm iteratively refines a candidate program using counterexamples generated during verification, ensuring that the synthesized program satisfies the desired properties. We tailor this algorithm to synthesize the linear policy for the programmatic runtime shield. Specifically, our algorithm initializes a linear policy using the \emph{Linear-Quadratic Regulator (LQR) algorithm}~\cite{anderson2007optimal} and iteratively refines it based on counterexamples identified by the \emph{maximal output admissible set}~\cite{gilbert1991linear}. We introduce our algorithm in the following paragraphs.


The LQR algorithm is a classic and effective method for designing optimal control policies for linear time-invariant systems~\cite{anderson2007optimal}. It is widely used in various fields, including robotics~\cite{mason2016balancing}, aerospace~\cite{yang2014quaternion}, and economics~\cite{yaghobipour2020solving}, due to its simplicity and effectiveness in managing multi-variable control systems. LQR derives control policies by minimizing a quadratic cost function, which is the sum of state and command costs. Specifically, given the state transition function $s_{t+1} = s_t + f(s_t, c_t) \times \Delta t$, where $f(s_t, c_t) = \mathbf{A}s_t + \mathbf{B}c_t$, the objective of the LQR is to determine the control input that minimizes the following quadratic cost function:
\begin{equation}
    J=\sum_{t=0}^{\infty}\left(s_t^\top \mathbf{Q} s_t+c_t^\top \mathbf{R} c_t\right)
\end{equation}
where the term $s_t^\top \mathbf{Q} s_t$ represents the state cost, and $c_t^\top \mathbf{R} c_t$ represents the command cost, with $s_t$ and $c_t$ being the state and command at time step $t$, respectively. The intuition behind these costs is that the state cost penalizes deviations of the system's current state from the desired state, encouraging the system to behave as expected, while the command cost penalizes the control commands, discouraging excessive actions and promoting smooth, efficient control that does not overly strain the system. The LQR algorithm computes the optimal control commands as a linear function of the current states, i.e.,
\begin{equation}
    c_t = -\mathbf{K}s_t
\end{equation}
where $\mathbf{K}$ is the matrix of the parameters in the linear function, and is computed  step by step to derive the discrete-time Algebraic Riccati Equation~\cite{anderson2007optimal, benner2020numerical}. Firstly, given the above quadratic cost function, we need to define a candidate value function $V_t(s_t)$, which is a quadratic function of the state $s_t$:
\begin{equation}
    V_t(s_t)=s_t^\top \mathbf{P} s_t
\end{equation}
where $\mathbf{P}$ is a positive semi-definite matrix that needs to be determined. We then apply the Bellman principle~\cite{bellman1957dynamic} to define the value function recursively:
\begin{equation}
    V_t(s_t) = \min_{c_t} \left[ s_t^\top \mathbf{Q} s_t + c_t^\top \mathbf{R} c_t + V_{t+1}(s_{t+1}) \right]
\end{equation}
Then we substitute the dynamics and the value function:
\begin{align}
    V_t(s_t) = \min_{c_t} \big[
        & s_t^\top \mathbf{Q} s_t + c_t^\top \mathbf{R} c_t + (s_t + \Delta t (\mathbf{A}s_t + \mathbf{B}c_t))^\top \mathbf{P} (s_t + \Delta t (\mathbf{A}s_t + \mathbf{B}c_t))
    \big]
\end{align}
Define the closed-form expansion:
\begin{align}
s_{t+1} &= s_t + \Delta t \mathbf{A} s_t + \Delta t \mathbf{B} c_t = (\mathbf{I} + \Delta t \mathbf{A}) s_t + \Delta t \mathbf{B} c_t
\end{align}
Let $\tilde{\mathbf{A}} = \mathbf{I} + \Delta t \mathbf{A}$ and $\tilde{\mathbf{B}} = \Delta t \mathbf{B}$. Then:
\begin{equation}
s_{t+1} = \tilde{\mathbf{A}} s_t + \tilde{\mathbf{B}} c_t
\end{equation}
The value function becomes:
\begin{align}
V_t(s_t) = \min_{c_t} \Big[
    & s_t^\top \mathbf{Q} s_t + c_t^\top \mathbf{R} c_t + (\tilde{\mathbf{A}} s_t + \tilde{\mathbf{B}} c_t)^\top \mathbf{P} (\tilde{\mathbf{A}} s_t + \tilde{\mathbf{B}} c_t)
\Big].
\end{align}
We then rearrange the above equation as:
\begin{align}
    V_t(s_t) = \min_{c_t} \Big[
        & s_t^\top (\mathbf{Q} + \tilde{\mathbf{A}}^\top \mathbf{P} \tilde{\mathbf{A}}) s_t
        + c_t^\top (\mathbf{R} + \tilde{\mathbf{B}}^\top \mathbf{P} \tilde{\mathbf{B}}) c_t + 2 s_t^\top \tilde{\mathbf{A}}^\top \mathbf{P} \tilde{\mathbf{B}} c_t
    \Big]
    \end{align}
Taking the derivative with respect to $c_t$ and setting to zero:
\begin{equation}
\frac{\partial V_t}{\partial c_t} = 2(\mathbf{R} + \tilde{\mathbf{B}}^\top \mathbf{P} \tilde{\mathbf{B}}) c_t + 2 \tilde{\mathbf{B}}^\top \mathbf{P} \tilde{\mathbf{A}} s_t = 0
\end{equation}
Then we solve $c_t$ which gives the optimal control command:
\begin{equation}
c_t = -\mathbf{K} s_t,
\quad \text{where} \quad
\mathbf{K} = (\mathbf{R} + \tilde{\mathbf{B}}^\top \mathbf{P} \tilde{\mathbf{B}})^{-1} \tilde{\mathbf{B}}^\top \mathbf{P} \tilde{\mathbf{A}}
\end{equation}
Plugging $c_t$ back into the value function, we obtain a new expression for $V_t(s_t)$, from which the updated matrix $\mathbf{P}_t$ can be derived recursively:
\begin{align}
\mathbf{P}_{t} =
\mathbf{Q} + \tilde{\mathbf{A}}^\top \mathbf{P}_{t+1} \tilde{\mathbf{A}}
- \tilde{\mathbf{A}}^\top \mathbf{P}_{t+1} \tilde{\mathbf{B}}
(\mathbf{R} + \tilde{\mathbf{B}}^\top \mathbf{P}_{t+1} \tilde{\mathbf{B}})^{-1}
\tilde{\mathbf{B}}^\top \mathbf{P}_{t+1} \tilde{\mathbf{A}}
\end{align}
This is the discrete-time algebraic Riccati recursion under Euler-discretized dynamics, which can be solved using value iteration~\cite{bellman1957dynamic, sutton1992reinforcement} or other numerical methods like Hamiltonian matrices~\cite{lancaster1995algebraic}.

Note that in our case we assume that the state cost matrix $\mathbf{Q}$ and the command cost matrix $\mathbf{R}$ are well-defined and provided by the user, and how to choose them is beyond the scope of this paper. We refer interested readers to~\cite{anderson2007optimal} for more details. Moreover, in our work the above calculation is done with an off-the-shelf solver, i.e., pyMPC~\cite{githubGitHubTobiaMarcuccipympc}.

After initializing the linear policy using LQR, our counterexample-guided inductive synthesis algorithm iteratively refines the policy until it is verified to be safe according to the safety specifications. To facilitate this process, we use the maximal output admissible set~\cite{gilbert1991linear} to identify counterexamples for further refinement. The maximal output admissible set is defined as the set of initial environment states for which the outputs of the linear policy are guaranteed to be safe commands. In other words, no matter which initial state within this set the environment starts from, the linear policy will never generate commands that could put the system in unsafe states at any time step. This set, denoted $\mathcal{X}$, can be defined as:
\begin{equation}
    \begin{aligned}
    &\mathcal{X}=\left\{s_0 \in \mathcal{S}_0 \mid \forall t \geq 0, s_t \in \neg \mathcal{S}_u\right.\\
    &\text { where } \left.s_{t+1}=s_t + f'\left(s_t, -\mathbf{K}s_t\right)\times \Delta t \right\}
    \end{aligned}
\end{equation}
where $\mathbf{K}$ represents the parameters of the linear policy initialized by the LQR algorithm, $f'$ is the inferred environment dynamics function obtained in Section~\ref{sec:state-transition}, $\mathcal{S}_0$ is the set of initial environment states, and $t$ is the time step. Here, $\neg \mathcal{S}_u$ is used to indicate the complement of the set of unsafe states, meaning it encompasses the set of safe states that the system should remain in, thereby serving as the safety specifications for the system.

The maximal output admissible set is computed by solving linear programming problems~\cite{gilbert1991linear}. Specifically, considering the safety specifications $\lnot\mathcal{S}_u$ defined by a set of linear inequalities, i.e., $\lnot\mathcal{S}_u = \{s \mid D_i s \leq d_i, i = 1, \dots, n\}$, where $D_i$ is a matrix of coefficients for each state variable, $d_i$ is a vector of constants, and $n$ is the number of constraints, these inequalities define the boundaries of the set of safe states. Our goal is to ensure that the outputs of the linear policy remain within this set at every time step. Starting from the environment states $s_1$ at time step $t = 1$, we check whether, given the dynamics function $f' = \mathbf{A}s_t + \mathbf{B}c_t$ and the linear policy $c_t = -\mathbf{K}s_t$, the next state $s{2} = s_1 + \mathbf{A}s_1 - \mathbf{B}\mathbf{K}s_1$ remains within the set of safe states $\neg\mathcal{S}_u$ (note at $t = 0$, verification is not needed since the next state is $s_1 = s_0$, which is already within $\neg\mathcal{S}_u$ and thus safe by definition). For each constraint, the corresponding linear programming problem is as follows:
\begin{equation}
    V(t=1, i)=\max _{s \in \neg\mathcal{S}_u}\left(D_i(s+\mathbf{A} s - \mathbf{B} \mathbf{K}s)-d_i\right)
\end{equation}
If for all $i$, $V(t = 1, i) < 0$, then the next state $s_2$ remains within the set of safe states $\neg\mathcal{S}_u$ for any state $s$ from $\neg\mathcal{S}_u$. According to the proved theorem in~\cite{gilbert1991linear}, this implies that all the future states of the system will also remain within the set of safe states $\neg\mathcal{S}_u$, provided with the initial state from $\neg\mathcal{S}_u$. Thus, we conclude the maximal output admissible set as $\mathcal{X} = \neg\mathcal{S}_u$. However, if one of the linear programming problems has $V(t = 1, i) \geq 0$, it indicates that there exists $s_1$ and $c_1$ such that $s_2$ could go outside $\neg\mathcal{S}_u$. In such a case, we then proceed to the next time step and check if $s_3$ remains within $\neg\mathcal{S}_u$, and continue this process. If, at any time step, all values of $V(t, i)$ are less than 0, we can conclude that all the future states from time step $t$ will stay within the set of safe states $\neg\mathcal{S}_u$, with the updated maximal output admissible set as the intersection of $\neg\mathcal{S}_u$, $A\neg\mathcal{S}_u$, $A^2\neg\mathcal{S}_u$, ..., up to the point where the iterative process stops. For ensuring $s_{t +1}$ stays within $\neg\mathcal{S}_u$, we need to check the following linear programming problem:
\begin{equation}
    V(t, i)=\max _{\substack{s \in \neg\mathcal{S}_u \cap A \neg\mathcal{S}_u \\ \cap \cdots \cap A^t \neg\mathcal{S}_u}}(D_i (s+\sum_{k=0}^{t-1}(\mathbf{A} s_k - \mathbf{B} \mathbf{K}s_k) \cdot k )-d_i)
\end{equation}
Although the above process sounds complex, our implementation leverages an off-the-shelf tool, i.e., pyMPC~\cite{githubGitHubTobiaMarcuccipympc}, which can efficiently solve the linear programming problems and compute the maximal output admissible sets. Note that in some extreme cases, there may be no time step $t \ge T$ where all $V(t, i)$ are less than 0, indicating a failure to compute the maximal output admissible set. In such scenarios, our method cannot derive runtime shields, leading to the termination of the synthesis process. However, in our experiments, this situation did not occur; the maximal output admissible set was successfully computed in all instances.

The utilization of the maximal output admissible set here is to identify counterexamples that show the linear policy is not universally safe. Intuitively, if the maximal output admissible set covers all initial states of the environment, the linear policy is theoretically verified to be safe, because as long as the system starts from an initial state within the maximal output admissible set, the linear policy will never generate unsafe commands at any time step, ensuring the system remains in safe states. However, if the maximal output admissible set does not cover all initial states, there exist counterexamples—specific initial states that cause the linear policy to produce unsafe commands at certain time steps. Formally, given the derived maximal output admissible set $\mathcal{X}$ and the set of initial environment states $\mathcal{S}_0$, the counterexamples can be defined as:
\begin{equation}
    \mathcal{Z} =\left\{s_0 \in \mathcal{S}_0 \mid s_0 \notin \mathcal{X}\right\}
\end{equation}
where $\mathcal{Z}$ is the set of counterexamples. If $\mathcal{Z}$ is not empty, it means that there exists an initial state $s_0$ in the environment such that the linear policy applied to $s_0$ may cause the system to enter unsafe states $\mathcal{S}_u$ at some future time steps, proving that the policy is not universally safe. Otherwise if $\mathcal{Z}  = \emptyset$, this implies that the linear policy is verified to be safe for all possible initial states $s_0$ in $\mathcal{S}_0$.

When counterexamples are found, we proceed to refine the linear policy until it is verified to be safe without any counterexamples. To achieve this, we propose using a gradient-based optimization algorithm, with the objective of maximizing the rewards provided by the environment while running the linear policy with the counterexamples. The intuition behind this objective is that the environment's reward function is typically designed to encourage the system to remain in safe states and penalize it for entering unsafe states. Therefore, by using the rewards as feedback, the linear policy is expected to increasingly generate safe commands after refinement. Additionally, this optimization algorithm differs from previous works~\cite{zhu2019inductive} that refine the linear policy to resemble the neural policy, because the neural policy is not guaranteed to be safe, and thus directly mimicking the neural policy may give inappropriate feedback for refining the linear policy, which could potentially lead the linear policy to adopt unsafe behaviors.

\begin{wrapfigure}{r}{0.5\textwidth}
    \begin{algorithm}[H]
     \small
    \SetKwInput{Input}{Require}
    \SetKwComment{Comment}{\textbf{**//**} }{}
    \SetCommentSty{upshape}
    \Input{$\mathbf{K}$, $\mathcal{Z} $, $f'$, $r$, $lr$, $T$}
    \For{$ce \in \mathcal{Z} $}{
    $\epsilon \gets \text{random\_perturbation}()$ \label{line:refine-linear-policy-1}\\
    $s_0 \gets ce$ \label{line:refine-linear-policy-2}\\
    \For{$\text{time step } t \gets 1$ \KwTo $T$}{
    $s_t = s_{t-1} + f'(s_{t-1}, -\mathbf{K}s_{t-1}) * t$ \label{line:refine-linear-policy-3}\\
    $c_t \gets -\mathbf{K}(s_t)$ \\
    $c^+_t \gets -(\mathbf{K} + \epsilon)s_t$, $c^-_t \gets -(\mathbf{K} - \epsilon)s_t$ \label{line:refine-linear-policy-4}\\
    $D_+ \gets - (c^+_t - c_t) + r(s_t, c^+_t)$ \label{line:refine-linear-policy-8}\\
    $D_- \gets - (c^-_t - c_t) + r(s_t, c^-_t)$ \label{line:refine-linear-policy-5}\\
    $\Delta \gets (D_+ - D_-) / (2 \times \epsilon)$ \label{line:refine-linear-policy-6}\\
    $\mathbf{K} \gets \mathbf{K} + lr \times \Delta$ \label{line:refine-linear-policy-7}
     }
     }
    \Return $\mathbf{K}$ \label{line:refine-linear-policy-9}
    \caption{Refine Linear Policy} \label{alg:refine-linear-policy}
    \end{algorithm}
    \end{wrapfigure}

We present the refinement algorithm in Algorithm~\ref{alg:refine-linear-policy}. The algorithm takes as input the linear policy parameters $\mathbf{K}$, the counterexample set $\mathcal{Z}$, the inferred environment dynamics $f'$, the environment's reward function $r$, the learning rate $lr$, and the total number of time steps $T$. For each counterexample $ce$ in the set $\mathcal{Z}$, the algorithm first randomly produces a small perturbation $\epsilon$ (line~\ref{line:refine-linear-policy-1}) to slightly alter the parameters of the linear policy $\mathbf{K}$ in the subsequent process. We then initialize the environment with the counterexample $ce$ (line~\ref{line:refine-linear-policy-2}) and run the system with the linear policy to evaluate the outputs of the linear policy, perturbed linear policies, and the rewards obtained from the environment (lines~\ref{line:refine-linear-policy-3}-\ref{line:refine-linear-policy-5}). With these outputs and rewards, we calculate their differences as an estimate of the gradients $\Delta$ for updating the linear policy (lines~\ref{line:refine-linear-policy-6}-\ref{line:refine-linear-policy-7}). The iteration continues until a given number of time steps is reached, and returns the refined parameters of the linear policy $\mathbf{K}$ (line~\ref{line:refine-linear-policy-9}). The algorithm intuitively refines the linear policy by maximizing the rewards obtained from the environment, which are designed to encourage the system to remain in safe states and penalize it for entering unsafe states. The small perturbation $\epsilon$ is used to estimate the gradients of the rewards with respect to the parameters of the linear policy~\cite{mania2018simple, zhu2019inductive}, which are then used to update the linear policy in the direction that maximizes the rewards.

\begin{wrapfigure}{r}{0.57\textwidth}
    \begin{algorithm}[H]
     \small
    \SetKwInput{Input}{Require}
    \SetKwComment{Comment}{\textbf{**//**} }{}
    \SetCommentSty{upshape}
    \Input{$f'$, $\mathbf{K}$, $\mathcal{S}_0$, $\lnot\mathcal{S}_u$, $r$, $lr$, $T$}
    $\mathcal{X} \gets \text{maximal\_output\_admissible\_set}(f', \mathbf{K}, \lnot \mathcal{S}_u)$ \label{line:counterexample-guided-synthesis-1}\\
    $\mathcal{Z} \gets \left\{s_0 \in \mathcal{S}_0 \mid s_0 \notin \mathcal{X}\right\}$ \label{line:counterexample-guided-synthesis-2}\\
    \While{$\mathcal{Z} \neq \emptyset$\label{line:counterexample-guided-synthesis-t}}{
    $\mathbf{K} \gets \text{refine\_linear\_policy}(\mathbf{K}, \mathcal{Z}, f', r, lr, T)$ \label{line:counterexample-guided-synthesis-3}\\
    $\mathcal{X} \gets \text{maximal\_output\_admissible\_set}(f', \mathbf{K}, \lnot \mathcal{S}_u)$\\
    $\mathcal{Z} \gets \left\{s_0 \in \mathcal{S}_0 \mid s_0 \notin \mathcal{X}\right\}$ \label{line:counterexample-guided-synthesis-4}\\
     }
    \Return $\mathbf{K}$ \label{line:counterexample-guided-synthesis-5}\\
    \caption{Counterexample-Guided Inductive Synthesis} \label{alg:counterexample-guided-synthesis}
    \end{algorithm}
    \end{wrapfigure}

We are prepared to present the overall counterexample-guided inductive synthesis algorithm, as shown in Algorithm~\ref{alg:counterexample-guided-synthesis}. The algorithm takes as input the inferred environment dynamics function $f'$, the initial linear policy $\mathbf{K}$, the set of initial environment states $\mathcal{S}_0$, the safety specifications of the environment $\lnot\mathcal{S}_u$, the reward function $r$, the learning rate $lr$, and the total number of time steps $T$. It first computes the maximal output admissible set $\mathcal{X}$ (line~\ref{line:counterexample-guided-synthesis-1}) and finds the set of counterexamples $\mathcal{Z}$ (line~\ref{line:counterexample-guided-synthesis-2}). The algorithm then iteratively refines the linear policy using the counterexamples and updates the maximal output admissible set, until no more counterexamples are found and the linear policy is verified to be safe against the safety specifications (lines~\ref{line:counterexample-guided-synthesis-3}-\ref{line:counterexample-guided-synthesis-4}). The algorithm finally returns the synthesized parameters $\mathbf{K}$ for the linear policy (line~\ref{line:counterexample-guided-synthesis-5}).

Note that the above algorithm is conditionally complete, meaning that if the maximal output admissible set is computable and the algorithm converges, it will eventually synthesize a linear policy that is verified to be safe, with all counterexamples eliminated. This follows simply from the fact that the termination condition of the iterative process is precisely when the computed MOAS fully covers the set of initial environment states $\mathcal{S}_0$, at which point no further counterexamples exist (see line~\ref{line:counterexample-guided-synthesis-t}).

\subsection{Switching Strategy}
\label{sec:switching-strategy-synthesis}

The switching strategy is the second main component of a programmatic runtime shield that we synthesize. In the program sketch, the switching strategy is a conditional statement that determines when to replace the potentially unsafe command with the safe command produced by the linear policy. The linear policies are typically less powerful than the neural policies in terms of maintaining stable systems (discussed in section~\ref{sec:discussion}). Therefore, invoking the linear policy too frequently to replace the commands from the neural policy may have a negative impact on the stability of the system, necessitating a switching strategy that minimizes unnecessary interventions to maintain permissiveness.

As shown in our program sketch (Section~\ref{sec:program-sketch}), the switching strategy has a threshold $\lambda$ to be synthesized, which is used to compare the difference between the command produced by the neural policy and the linear policy. If the difference is greater than the threshold $\lambda$, the linear policy is invoked to replace the command from the neural policy. A smaller threshold $\lambda$ makes the switching strategy more sensitive to the difference, leading to more frequent interventions of the linear policy, while a larger threshold $\lambda$ makes the switching strategy too permissive and may fail to correct the unsafe commands from the neural policy. Thus, the threshold $\lambda$ should be carefully chosen to balance the trade-off, which is essentially an optimization problem.

We propose using a Bayesian optimization algorithm to solve this problem. Bayesian optimization is a powerful tool for optimizing black-box functions that are expensive to evaluate~\cite{10.1145/3520304.3533654,snoek2012practical}, which is widely used in various software-related domain such as compiler optimization~\cite{10.1145/3623278.3624770,9401979}.
. It maintains a surrogate model, which approximates the real objective function but at a much lower computational cost. During the optimization process, candidate solutions are evaluated using the surrogate model, and the real objective function is evaluated only when a promising solution is identified. This approach significantly reduces the cost of evaluating candidate solutions, making it particularly well-suited for computationally expensive optimization problems.

In our case, we aim to optimize the threshold $\lambda$ of the switching strategy to minimize the number of violations of the safety specifications while maximizing the ratio of necessary interventions to total interventions. Formally, denoting the number of violations as $\mathcal{V}$, the number of necessary interventions as $\mathcal{V}^*$, and the total number of interventions as $\mathcal{I}$, the objective function can be formulated as:
\begin{equation}
    \label{eq:objective-function}
    \arg \min _\lambda\left(\log (\mathcal{V}+1)-\frac{\mathcal{V}^*}{\mathcal{I}+1}\right)
\end{equation}
where the $log$ function can make the optimization more sensitive to the small number of violations, ensuring that the algorithm focuses on minimizing the number of violations. The adding of 1 in the logarithm and the denominator is to avoid undefined values when the number of violations or interventions is zero. Intuitively, the ideal threshold $\lambda$ derived by the optimization algorithm with this objective function should minimize the number of violations to zero while maximizing the ratio of necessary interventions to total interventions, ensuring that the system remains safe with minimal interventions from the linear policy.

Note that this objective function is expensive to evaluate, because $\mathcal{V}$, $\mathcal{V}^*$, and $\mathcal{I}$ are unknown and cannot be directly computed using $\lambda$. Instead, they can only be obtained by running the environment with the linear policy and the neural policy to count the number of violations and interventions. Therefore, in our Bayesian optimization algorithm, we use the Gaussian process (GP) as a surrogate model of the objective function, which is a widely-used model in Bayesian optimization~\cite{binois2022survey}. The GP maintains a Gaussian distribution over the objective function to approximate the true function, which requires only a few evaluations of the objective function to guide the search for the optimal solution. Specifically,
the GP can be modeled as:
\begin{equation}
    obj(\lambda) \sim \mathcal{GP}(\mu(\lambda), k(\lambda, \lambda'))
\end{equation}
where $obj(\lambda)$ is the objective function that we aim to optimize, $\mu(\lambda)$ and  $k(\lambda, \lambda')$ are called the mean function and the kernel function, respectively. We follow the implementation of the Gaussian process in the scikit-learn library~\cite{scikitoptimizeSkoptgp} to construct the model with the zero-mean function and the Matern kernel function~\cite{genton2001classes}. With each evaluation of the objective function, the GP model updates its parameters to better approximate the true function. Once the GP model is constructed and updated, it can predict the output of the objective function at a new input $\lambda$ as $\mu(\lambda)$. The Bayesian optimization algorithm iteratively refines the GP model and selects the next candidate solution to evaluate based on the predicted output of the GP, which guides the search for the optimal solution.

In addition, the acquisition function of the Gaussian process is used to select the next candidate solutions over the current best solution. We follow~\cite{chen2021efficient} to use expected improvement~\cite{benassi2011robust} as the acquisition function, which measures the expected improvement of the surrogate model over the current best solution and selects the candidate solution that maximizes the expected improvement. Formally, the expected improvement is defined as:
\begin{equation}
    \text{EI}(\lambda) = \mathbb{E}[\max(0, \mu(\lambda)^* - \mu(\lambda))]
\end{equation}
where $\mu(\lambda)^*$ is the current best solution, $\mu(\lambda)$ is the predicted output of the Gaussian process at input $\lambda$, and $\mathbb{E}$ is the expectation. With the acquisition function, in each iteration, the Bayesian optimization algorithm selects the candidate solution that maximizes the expected improvement, evaluates the objective function at the candidate solution, and updates the GP model to better approximate the true function. The iteration continues until the number of iterations is reached, and the algorithm returns the synthesized threshold $\lambda$.



%% file: sections/results.tex
\section{Evaluation}
\label{sec:results}

This section presents the evaluation of \tool.
Our experiments aim to answer the following research questions:
\begin{itemize}[leftmargin=*]
    \item \textbf{RQ1 (Effectiveness):} Can the shields synthesized by \tool effectively maintain the safety of the neural policy?
    \item \textbf{RQ2 (Efficiency):} What is the performance overhead of \tool's shields in terms of time and memory?
    \item \textbf{RQ3 (Permissiveness):} How permissive are \tool's shields in terms of imposing shield interventions as few as possible?
    \item \textbf{RQ4 (Scalability):} How scalable is the shield synthesis process of \tool?
\end{itemize}

\subsection{Experimental Setup}
\label{sec:Experimental_Setup}

\paragraph{Experimental Subjects}
To comprehensively evaluate \tool, we select eight control systems that are widely used in prior work~\cite{zhu2019inductive,10.1145/3715105}, covering a range of real-world scenarios including autonomous driving (Self-Driving and 4-Car Platoon), aerial drone control (Quadcopter), and dynamic balancing (CartPole and Pendulum). Table~\ref{tab:benchmarks} summarizes each control system along with its associated safety properties. Specifically,
\begin{itemize}
  \item {\sf Quadcopter}: This system stabilizes a quadcopter in flight by ensuring the pitch angle $\eta_1$ and yaw angle $\eta_2$ remain within $\pm \pi/2$ radians.
  \item {\sf CartPole-v1} and {\sf CartPole-v2}: Both systems involve a pole mounted on a cart that moves along a track. Safety is defined by the following constraints: the cart position $|\delta| < 0.3$ m, cart velocity $|\delta'| < 0.5$ m/s, pole angle $|\eta| < 30^\circ$, and pole angular velocity $|\eta'| < 0.5$ rad/s. The two variants differ in the complexity of neural policies: CartPole-v1 uses a neural policy with 0.7M neurons, while CartPole-v2 employs a larger policy with 7M neurons.
  \item {\sf Pendulum-v1} and {\sf Pendulum-v2}: These two systems aim to balance a pendulum in an upright position. Safety is maintained by restricting the angle $\eta$ to within $30^\circ$ and angular velocity $\omega$ to within $30^\circ$/s. The main difference lies in the pendulum length: 1.0 m for Pendulum-v1 and 1.5 m for Pendulum-v2.
  \item {\sf Self-Driving-v1} and {\sf Self-Driving-v2}: These systems simulate autonomous driving at constant speed, where safety requires the car's heading angle $\eta$ to stay within $90^\circ$, and the lateral distance $d$ from the centerline to be less than 2.0 m for v1 and 1.5 m for v2, ensuring the vehicle does not veer into roadside canals.
  \item {\sf 4-Car Platoon}: This system models four vehicles maintaining safe distances. Each variable in the above specification is the distance between two consecutive vehicles in the platoon, and safety is assured if they are below certain thresholds.
\end{itemize}

\begin{table*}[t!]
  \small
  \caption{Control systems and their safety specifications. The detailed description of these systems and specifications are in Section~\ref{sec:Experimental_Setup}. The table also shows the number of inputs, outputs, and neural networks sizes of neural policies.}
  \renewcommand{\familydefault}{\sfdefault}\normalfont
  \label{tab:benchmarks}
  \centering
  \scalebox{0.95}{
  \begin{tabular}{c|c|c|c|c}
  \hline
  \hline
  \multirow{2}{*}{Control System} & \multirow{2}{*}{ Safety Specifications} & \multirow{2}{*}{\begin{tabular}[c]{@{}c@{}}\# Inputs/\\ \# Outputs\end{tabular}} & \multicolumn{2}{c}{Neural Policy Size} \\
  \cline{4-5}
   &  &   & \#Layers & \#Neurons (M, a million) \\ \hline \hline
  Quadcopter  & $ \vert\eta_1\vert <\pi/2 \wedge\vert \eta_2 \vert <\pi/2 $   & 2/1 & 2 & 0.4 M \\ \hline
  CartPole-v1   & \makecell{
      $|\delta| < 0.3 \;\wedge\; |\delta'| < 0.5 \;\wedge$ \\
      $|\eta| < 30^\circ \;\wedge\; |\eta'| < 0.5$
      } & 4/1 & 3 & 0.7 M \\ \hline
  CartPole-v2   & \makecell{
      $|\delta| < 0.3 \;\wedge\; |\delta'| < 0.5 \;\wedge$ \\
      $|\eta| < 30^\circ \;\wedge\; |\eta'| < 0.5$
      } & 4/1 & 3 & 7 M \\ \hline
  Pendulum-v1   & $ \vert\eta\vert <30^{\circ} \wedge\vert \omega \vert <30^{\circ} $  & 2/1 & 3 & 7.0 M \\
  \hline
  Pendulum-v2   & $ \vert\eta\vert <30^{\circ} \wedge\vert \omega \vert <30^{\circ} $  & 2/1 & 3 & 7.0 M \\ \hline
  Self-Driving-v1 & $ \vert\eta\vert <90^{\circ} \wedge\vert d\vert < 2.0 $   & 2/1 & 3& 0.7 M   \\ \hline
  Self-Driving-v2 & $ \vert\eta\vert <90^{\circ} \wedge\vert d\vert < 1.5 $   & 2/1 & 3& 0.7 M   \\ \hline
  4-Car Platoon & \makecell{
      $ \vert a \vert <= 2.0 \wedge \vert b \vert <= 0.5 \wedge$ \\
      $ \vert c \vert <= 0.35 \;\wedge\; \vert d \vert <= 0.5 \;\wedge$\\
      $ \vert e \vert <= 1.0 \;\wedge\; \vert f \vert <= 0.5 \;\wedge \vert g \vert <= 1.0$
      } & 7/4 & 3& 7.0 M   \\ \hline \hline
  \end{tabular}
  }
\end{table*}

While these systems may appear simple, they are still representative of practical control systems that are difficult to control and maintain safety, such as Segway transporters~\cite{Segway}, camera drones~\cite{DJI}, and autonomous driving cars~\cite{Udacity}. Therefore, we believe that these control systems are appropriate for evaluating \tool and the baseline.

We follow Zhu et al.~\cite{zhu2019inductive} to train the neural policies for the eight control systems using the Deep Deterministic Policy Gradient (DDPG)~\cite{LillicrapHPHETS15}, a popular DRL algorithm for continuous control tasks. All hyperparameters are set to the default values from Zhu et al.'s official implementation. Note that our method is agnostic to the choice of DRL algorithms; other algorithms could also be applied. We choose DDPG because it is widely used in the literature~\cite{wang2022deep} and aligns with the experimental setup of our baseline. We use the same neural network architecture as \cite{zhu2019inductive} for each control system, which consists of 2 or 3 fully connected layers. The neural policies range in size from 0.4M to 7M neurons, as detailed in Table~\ref{tab:benchmarks}. Notably, these sizes are significantly larger than those in previous work~\cite{zhang2022falsifai, tran2020nnv}, where policies typically contain no more than 400 neurons.

\paragraph{Experimental Settings and Metrics}
For each control system, we evaluate the performance of \tool and the baseline \baseline~\cite{zhu2019inductive} by synthesizing programmatic runtime shields for the neural policies. Note that \baseline's shields apply barrier certificates with high-degree polynomials while our \tool synthesizes shields with lightweight linear policies. We follow \baseline's original implementation to set the degree of the polynomials for the barrier certificates to 4, which is the value recommended by the authors. We also tried setting the degree to 1, i.e., making the barrier certificates linear, but \baseline's synthesis process failed to converge and cannot find shields that can guarantee safety. Therefore, we use the default setting of \baseline to ensure a fair comparison. Moreover, the hyperparameter for inferring environment dynamics, i.e., $\epsilon$ in Algorithm~\ref{alg:linearization}, is set to 1.49e-10, following the default configuration used in widely-used scientific computing libraries such as SciPy~\cite{scipyApprox_fprimex2014}.

We answer the research questions by comparing the performance of \tool and baseline in terms of the following metrics. The effectiveness of each tool is measured by the number of safety violations found in the systems using the shields synthesized by each tool. The efficiency of the shields of each tool is measured by their time and memory overhead on the systems. The permissiveness of each tool is measured by the number of shield interventions required by each tool's shields, as well as the ratio of the number of necessary shield interventions to the total number of interventions. The scalability of each tool is measured by the time required to synthesize a shield for the neural policy.

Additionally, to mitigate the impact of randomness, we repeat each experiment 10 times and report the average results along with their statistical significance and effect size as suggested by Arcuri and Briand~\cite{arcuri2011practical}. We use the Mann-Whitney U test~\cite{mann1947test} to measure the statistical significance and the Vargha-Delaney statistic $\hat{A}_{12}$~\cite{vargha2000critique} to measure the effect size. All the experiments are conducted on an Ubuntu 18.04 server equipped with an Intel Xeon E5-2698 CPU and 504 GB RAM.

\subsection{RQ1: Effectiveness of \tool}

\begin{table}[t!]
    \centering
    \caption{Effectiveness of \tool compared to the baseline. The table shows the number of safety violations found by \tool and the baseline \baseline.}
    \renewcommand{\familydefault}{\sfdefault}\normalfont
    \label{tab:RQ1}
    \begin{tabular}{c|c|c|c}
    \hline\hline
    \multirow{2}{*}{System} & \multicolumn{3}{c}{\# Safety Violations} \\ \cline{2-4}
    & \multicolumn{1}{c}{Neural Policy} & \multicolumn{1}{c|}{w/ \baseline} & \multicolumn{1}{c}{w/ \tool} \\ \hline\hline
    Quadcopter & 182/1000 & {\bf 0/1000} & {\bf 0/1000} \\ \hline
    Pendulum-v1 & 41/1000 & {\bf 0/1000} & {\bf 0/1000} \\ \hline
    Pendulum-v2 & 166/1000 & {\bf 0/1000} & {\bf 0/1000} \\ \hline
    Cart-Pole-v1 & 47/1000 & {\bf 0/1000} & {\bf 0/1000} \\ \hline
    Cart-Pole-v2 & 96/1000 & {\bf 0/1000} & {\bf 0/1000} \\ \hline
    Self-Driving-v1 & 35/1000 & {\bf 0/1000} & {\bf 0/1000}  \\ \hline
    Self-Driving-v2 & 60/1000 & {\bf 0/1000} & {\bf 0/1000} \\ \hline
    4-Car Platoon & 7/1000 & {\bf 0/1000} & {\bf 0/1000} \\ \hline\hline
    Average & 79/1000 (7.9\% Unsafe) & {\bf 0/1000 (100\% Safe)} & {\bf 0/1000 (100\% Safe)} \\
    \hline\hline
    \end{tabular}
\end{table}

This RQ evaluates the effectiveness of \tool in ensuring the safety of neural policies across eight control systems. Following Zhu et al.~\cite{zhu2019inductive}, we run each neural policy for 1,000 episodes, each consisting of 5,000 time steps with randomized initial states. We compare the number of safety violations under three settings: using the neural policy alone, using the baseline's shield, and using the shield synthesized by \tool, as summarized in Table~\ref{tab:RQ1}.

The results show that unshielded neural policies violate safety specifications in all systems, with an average violation rate of 7.9\%, corresponding to 79 violations per 1,000 runs. This highlights the inherent risks of deploying neural policies without formal safety enforcement, especially in stochastic or dynamic environments where policy behaviors may deviate from expected safe trajectories.

When either \tool or the baseline's shield is applied, all safety violations are eliminated across all systems—that is, the number of violations drops to 0—demonstrating that both methods effectively intercept and correct unsafe commands at runtime. The fact that \tool achieves the same level of safety as the baseline confirms its ability to synthesize shields of equivalent effectiveness. These results also suggest that \tool is generalizable, successfully enforcing safety across a diverse range of control systems with varying specifications and neural policy sizes, thereby underscoring its potential for deployment in real-world safety-critical applications.

\ans{\textbf{RQ1 Takeaway.}\hspace{6pt} \tool can effectively ensure the safety of the neural policies in all control systems by correcting all the unsafe commands and letting the systems operate without any violations in safety specifications.}

\subsection{RQ2: Overhead of \tool}

A programmatic runtime shield should be able to respond quickly and consume minimal resources, so we compare the time and memory overhead of the shields synthesized by \tool and \baseline across the eight control systems.

\begin{table}[t!]
    \small
    \centering
    \caption{Overhead of \tool compared to the baseline. The table shows the time and memory overhead of \tool compared to the baseline. The last two columns show the statistical significance and effect size of the overhead.}
    \renewcommand{\familydefault}{\sfdefault}\normalfont
    \label{tab:RQ2}
    \scalebox{0.87}{
    \begin{tabular}{c|c|c|c|c|c|c|c|c}
    \hline\hline
    \multirow{2}{*}{Control System}
      & \multicolumn{4}{c}{Time Overhead (s)}
      & \multicolumn{4}{c}{Memory Overhead (MB)} \\ \cline{2-9}
    & \baseline & \tool
      & {\it p}-value& {\it \^A}\textsubscript{12}
      & \baseline& \tool
      & {\it p}-value & {\it \^A}\textsubscript{12}\\ \hline\hline
    Quadcopter      & 1.38s & {\bf 0.60s (2.3$\times$)}
      & 4.7e-4$<$0.01 & {\it L} (1.00)
      & 1.98 MB & {\bf 0.53 MB (3.7$\times$)}
      & 0.03$<$0.05 & {\it L} (0.73) \\ \hline
    Pendulum-v1     & 2.07s & {\bf 0.71s (2.9$\times$)}
      & 8.8e-5$<$0.01 & {\it L} (1.00)
      & 4.05 MB & {\bf 0.76 MB (5.3$\times$)}
      & 2.2e-4$<$0.01 & {\it L} (0.97) \\ \hline
    Pendulum-v2     & 2.45s & {\bf 1.12s (2.2$\times$)}
      & 9.5e-5$<$0.01 & {\it L} (1.00)
      & 6.70 MB & {\bf 0.76 MB (8.8$\times$)}
      & 9.6e-5$<$0.01 & {\it L} (1.00) \\ \hline
    Cart-Pole-v1    & 2.81s & {\bf 1.31s (2.1$\times$)}
      & 8.5e-4$<$0.01 & {\it L} (0.80)
      & 1.49 MB & {\bf 0.58 MB (2.6$\times$)}
      & 2.9e-4$<$0.01 & {\it L} (0.96) \\ \hline
    Cart-Pole-v2    & 1.93s & {\bf 1.48s (1.3$\times$)}
      & 3.3e-4$<$0.01 & {\it L} (0.87)
      & 1.48 MB & {\bf 0.79 MB (1.9$\times$)}
      & 7.8e-4$<$0.01 & {\it L} (1.00) \\ \hline
    Self-Driving-v1 & 2.45s & {\bf 1.82s (1.3$\times$)}
      & 6.1e-4$<$0.01 & {\it L} (0.80)
      & 3.13 MB & {\bf 0.57 MB (5.5$\times$)}
      & 6.0e-3$<$0.01 & {\it L} (1.00) \\ \hline
    Self-Driving-v2 & 3.48s & {\bf 1.04s (3.3$\times$)}
      & 6.6e-4$<$0.01 & {\it L} (1.00)
      & 4.76 MB & {\bf 0.55 MB (8.7$\times$)}
      & 7.2e-4$<$0.01 & {\it L} (1.00) \\ \hline
    4-Car Platoon   & 5.47s & {\bf 2.16s (2.5$\times$)}
      & 5.7e-5$<$0.01 & {\it L} (1.00)
      & 3.13 MB & {\bf 2.28 MB (1.4$\times$)}
      & 3.0e-3$<$0.01 & {\it L} (1.00) \\ \hline\hline
    Average         & 2.76s & {\bf 1.28s (2.2$\times$)} & --- & ---
      & 3.34 MB & {\bf 0.85 MB (3.9$\times$)} & --- & --- \\ \hline\hline
    \end{tabular}
    }
\end{table}

\paragraph{Time Overhead}
We measure time overhead as the additional operating time introduced by deployed shields during system execution. Table~\ref{tab:RQ2} presents the time overhead of \tool compared to the baseline. On average, the programmatic runtime shield synthesized by \tool incurs a time overhead of just 1.28 seconds, which is 2.2$\times$ lower than the 2.76 seconds incurred by the baseline. This consistent performance across all systems indicates that \tool generates lightweight shields with reduced runtime latency. This efficiency is largely due to \tool's use of linear templates for shield synthesis, which are simpler and faster to evaluate than the baseline's higher-order polynomial expressions. In real-time control systems, such reductions in latency can significantly enhance system responsiveness and reduce the risk of actuation delays.

To ensure the robustness of our findings and account for randomness, we apply the Mann-Whitney U test~\cite{mann1947test}, a non-parametric statistical test for comparing two independent groups. We use this test to evaluate whether the difference in time overhead between the two methods is statistically significant across 10 experimental runs. As shown in Table~\ref{tab:RQ2}, the time overhead of \tool is statistically significantly lower than that of \baseline, with all p-values well below 0.01, indicating statistical significance at the 99\% confidence level and a very low probability (less than 1\%) of being due to random chance.

We also report the Vargha-Delaney statistic $\hat{A}_{12}$~\cite{vargha2000critique} to quantify the effect size. This non-parametric measure gives the probability that a randomly chosen value from one group is higher or lower than one from another group. It is commonly used to evaluate search-based testing methods due to their randomness~\cite{meng2022linear,olsthoorn2020generating}. According to Vargha and Delaney~\cite{vargha2000critique}, $\hat{A}_{12}$ values greater than 0.71 (or less than 0.29) indicate a ``large'' effect size. Values between 0.64 and 0.71 (or between 0.29 and 0.36) correspond to a ``medium'' effect size, while values between 0.36 and 0.64 suggest a ``negligible'' effect size. As shown in Table~\ref{tab:RQ2}, \tool outperforms \baseline with a large effect size across all eight control systems, with all $\hat{A}_{12}$ values exceeding 0.80. This reflects a strong and consistent advantage of \tool over \baseline -- in at least 80\% of the experiments, \tool incurs lower time overhead than \baseline.

\paragraph{Memory Overhead}
We then compare the memory overhead of the shields synthesized by \tool and the baseline after deploying them into the eight control systems. We measure the memory overhead as the additional memory required by the deployed shields in the systems. Table~\ref{tab:RQ2} shows the memory overhead of \tool compared to the baseline. On average, the programmatic runtime shield synthesized by \tool occupies only 0.85 MB of memory, while the baseline's shields occupy 3.34 of MB memory, which suggests that \tool greatly reduces the memory overhead by 3.9$\times$ compared to the baseline.

This considerable saving arises because \tool represents the shield as a compact linear function and a threshold-based switching rule, requiring far fewer coefficients and structural components than the baseline's more complex symbolic programs. The reduction is especially notable in systems such as Pendulum and Self-Driving, where \tool achieves up to 8.8$\times$ lower memory usage. This compactness makes \tool particularly attractive for resource-constrained settings, such as embedded systems, drones, or edge computing devices.

Similarly, we run each experiment 10 times and use the Mann-Whitney U test to validate the statistical significance and the Vargha-Delaney statistic $\hat{A}_{12}$ to measure the effect size of improvements. Our results show that \tool significantly outperforms \baseline on all eight control systems at a 99\% confidence level in terms of memory overhead, except for the Quadcopter system, where the improvement is still statistically significant at the 95\% confidence level. The effect size of the memory overhead is also large for all control systems, with $\hat{A}_{12}$ values exceeding 0.73, indicating that \tool's shields are consistently more efficient than those of the baseline across all systems with low probability of being due to random chance.

\ans{\textbf{RQ2 Takeaway.}\hspace{6pt} The shields synthesized by \tool significantly outperform those of the baseline in terms of time and memory overhead on all control systems. Specifically, \tool's shields incur only 1.28s time overhead and occupy only 0.85 MB memory on average, which are 2.2$\times$ and 3.9$\times$ improvements over the baseline, respectively.}

\subsection{RQ3: Permissiveness of \tool}

\begin{table}[t!]
    \caption{Permissiveness of \tool compared to the baseline. The table shows the number of shield interventions required by \tool and the baseline. The last two columns show the statistical significance and effect size of the number of shield interventions.}
    \renewcommand{\familydefault}{\sfdefault}\normalfont
    \label{tab:RQ3}
    \centering
    \begin{tabular}{c|c|c|c|c}
    \hline\hline
    \multirow{2}{*}{Control System} & \multicolumn{4}{c}{\# Shield Interventions} \\ \cline{2-5}
    & \baseline & \tool & {\it p}-value& {\it \^A}\textsubscript{12} \\ \hline\hline
    Quadcopter      & 296   & {\bf 232 (1.3$\times$)} & 8.3e-3 $<$ 0.01 & {\it L} (0.92) \\ \hline
    Pendulum-v1     & 50    & {\bf 45 (1.1$\times$)}  & 8.6e-5 $<$ 0.01 & {\it L} (1.00) \\ \hline
    Pendulum-v2     & 211   & {\bf 120 (1.8$\times$)} & 6.0e-4 $<$ 0.01 & {\it L} (1.00) \\ \hline
    Cart-Pole-v1    & 1941  & {\bf 1133 (1.7$\times$)}& 9.8e-5 $<$ 0.01 & {\it L} (1.00) \\ \hline
    Cart-Pole-v2    & 826   & {\bf 690 (1.2$\times$)} & 9.0e-4 $<$ 0.01 & {\it L} (1.00) \\ \hline
    Self-Driving-v1 & 147   & {\bf 127 (1.2$\times$)} & 6.1e-3 $<$ 0.01 & {\it L} (1.00) \\ \hline
    Self-Driving-v2 & 292   & {\bf 139 (2.1$\times$)} & 5.8e-5 $<$ 0.01 & {\it L} (1.00) \\ \hline
    4-Car Platoon   & 18    & {\bf 17 (1.1$\times$)}  & 8.3e-3 $<$ 0.01 & {\it L} (0.77) \\ \hline\hline
    Average         & 473 & {\bf 313 (1.5$\times$)} & ---             & ---          \\ \hline\hline
    \end{tabular}
\end{table}

We first compare the permissiveness of the shields synthesized by \tool and \baseline in terms of the number of shield interventions over 1,000 runs (each run has 5,000 time steps) on the eight control systems. Table~\ref{tab:RQ3} shows the number of shield interventions imposed by the shields of each tool. On average, \tool requires only 313 shield interventions to ensure safety, which is 1.5$\times$ fewer than the 473 shield interventions required by the baseline. This consistent reduction suggests that \tool not only enforces safety but does so with greater permissiveness, allowing the neural policy to operate freely when safe. This improvement is primarily attributed to our switching strategy synthesized via Bayesian optimization, which identifies an optimal threshold that avoids unnecessary or overly conservative interventions. In contrast, the baseline tends to trigger interventions more frequently, including when they are not necessary to avoid safety violations.

Similar to the analysis in RQ2, we have conducted 10 independent runs and applied the Mann-Whitney U test to assess statistical significance. As shown in the last two columns of Table~\ref{tab:RQ3}, all eight systems show $p$-values below 0.01, indicating the difference is unlikely due to chance. Furthermore, the Vargha-Delaney $\hat{A}_{12}$ effect sizes are large in most systems, reaching 1.00 in five cases, confirming that \tool's improvement is both statistically and practically significant.

We also compare the ratio of necessary shield interventions to total interventions, as shown in Figure~\ref{fig:RQ3}. A necessary intervention is one that prevents a violation of the safety specifications, while an unnecessary one occurs when no violation would have happened even without intervention. Notably, the number of necessary interventions does not always equal the number of safety violations observed under the original (unshielded) neural policy, as reported in RQ1. This discrepancy arises because, once a shield intervenes, it modifies the system's trajectory by altering the original control command. While this new trajectory may avoid some original violations, it can also introduce new potential violations at other time steps that would not have occurred without shielding. Across all systems, \tool's shields consistently impose a higher ratio of necessary interventions, averaging 10.2\% more than the baseline. This indicates that \tool's shields are more precise, intervening only when safety is genuinely at risk, rather than interrupting benign commands from neural policies. Such precision is crucial for maintaining system stability while ensuring safety. A permissive shield improves overall stability, further reduces actuation delays, and increases trust in the policy during real-world deployment.

\input{figures/RQ2.tex}

\ans{\textbf{RQ3 Takeaway.}\hspace{6pt}\tool's shields significantly outperform those of baseline in terms of permissiveness on all control systems. Specifically, \tool requires only 384 shield interventions on average, which is a 1.5$\times$ improvement over the baseline. Furthermore, \tool's shields have a 10.2\% higher ratio of necessary shield interventions to total number of interventions than the baseline's shields.}

\begin{table}[t!]
    \centering
   \caption{Time cost of \tool compared to \baseline. The table shows the time required to synthesize a shield for the neural policies. The last column shows the statistical significance and effect size of the time cost.}
   \renewcommand{\familydefault}{\sfdefault}\normalfont
   \label{tab:RQ4}
   \begin{tabular}{c|c|c|c|c}
   \hline\hline
   \multirow{2}{*}{ Control System} & \multicolumn{4}{c}{Time Cost of Shield Synthesis}\\ \cline{2-5}
   & \multicolumn{1}{c|}{\baseline} & \multicolumn{1}{c|}{\tool} & \multicolumn{1}{c|}{ {\it p}-value} & {\it \^A}\textsubscript{12}\\ \hline\hline
    Quadcopter & 28s & {\bf 14s (2.0$\times$)} & 2.2e-4 \textless \ 0.01 & {\it L} (0.97)\\ \hline
    Pendulum-v1 & 439s & {\bf 85s (5.2$\times$)} & 1.8e-4 \textless \ 0.01 & {\it L} (1.00) \\ \hline
    Pendulum-v2 & 676s & {\bf 254s (2.7$\times$)} & 8.8e-3 \textless \ 0.01 & {\it L} (1.00) \\ \hline
    Cart-Pole-v1 & 395s & {\bf 118s (3.3$\times$)} & 9.3e-5 \textless \ 0.01 & {\it L} (1.00) \\ \hline
    Cart-Pole-v2 & 503s & {\bf 222s (2.3$\times$)} & 8.4e-3 \textless \ 0.01 & {\it L} (1.00) \\ \hline
    Self-Driving-v1 & 281s & {\bf 43s (6.5$\times$)} & 9.5e-5 \textless \ 0.01 & {\it L} (1.00) \\ \hline
    Self-Driving-v2 & 226s & {\bf 46s (4.9$\times$)} & 1.6e-4 \textless \ 0.01 & {\it L} (1.00) \\ \hline
    4-Car Platoon & 149s & {\bf 117s (1.3$\times$)} & 1.1e-3 \textless \ 0.01 & {\it L} (0.89) \\ \hline\hline
    Average & 337s & {\bf 112s (3.0$\times$)} & ---& ---\\
   \hline\hline
   \end{tabular}
   \end{table}

\input{figures/RQ3.tex}

\subsection{RQ4: Scalability of \tool}
In the final research question, we evaluate the scalability of \tool by comparing the time required to synthesize a shield for a neural policy. Table~\ref{tab:RQ4} presents the synthesis time of \tool and the baseline. On average, \tool completes shield synthesis in 112 seconds, achieving a 3.0$\times$ reduction in time compared to the baseline, which requires 337 seconds. This demonstrates that \tool is significantly more efficient across all control systems. The speedup is primarily due to \tool's formulation of shield synthesis as a sketch-based search over linear templates, which reduces the search space and enables faster convergence. In contrast, VRL employs high-degree polynomial templates that demand more costly constraint solving, making it less scalable as system complexity increases.

To check for statistical significance, we perform 10 independent runs per system and apply the Mann-Whitney U test. All $p$-values are well below 0.01, confirming the improvements are statistically significant. The Vargha-Delaney $\hat{A}_{12}$ values also indicate large effect sizes across all systems, showing that the improvements are substantial in practice.

Beyond Table~\ref{tab:RQ4}, we also draw a line chart to compare the time required to synthesize a shield for neural policies of different sizes, shown in Figure~\ref{fig:RQ4}. As Zhu et al.~\cite{zhu2019inductive} and Shi et al.~\cite{10.1145/3715105} provide neural policies of different sizes for the eight control systems, we use these neural policies to evaluate the scalability of \tool.
We divide the neural policies into three groups based on their sizes: small (0.4M neurons), medium (0.7M neurons), and large (7M neurons). Note that such sizes are sufficiently large for neural policies, especially when contrasted with policies used in previous work such as~\cite{zhang2022falsifai, tran2020nnv}, which typically contain no more than 400 neurons.

Figure~\ref{fig:RQ4} shows that \tool's synthesis time remains relatively stable as the size of the neural policies increases, whereas the baseline's synthesis time grows significantly. This result demonstrates that \tool scales more effectively with policy size and handles increasing complexity gracefully. In contrast, the baseline's time increases sharply, likely due to the greater burden of constraint solving in more complex settings. This scalability is crucial in practice, as modern applications in robotics, autonomous driving, and cyber-physical systems increasingly rely on large neural policies. \tool's ability to handle such policies without a substantial increase in synthesis time makes it a more practical solution for real-world deployment.

\ans{\textbf{RQ4 Takeaway}\hspace{6pt}\tool consistently shows better scalability than the baseline on all control systems. Specifically, \tool requires 112s on average for synthesizing a shield for a neural policy, which is 3.0$\times$ faster than the baseline. Furthermore, \tool's synthesis time does not significantly increase with the size of the neural policies.}

%% file: figures/RQ2.tex
\tikzdeclarepattern{
  name=mylines,
  parameters={
      \pgfkeysvalueof{/pgf/pattern keys/size},
      \pgfkeysvalueof{/pgf/pattern keys/angle},
      \pgfkeysvalueof{/pgf/pattern keys/line width},
  },
  bounding box={
    (0,-0.5*\pgfkeysvalueof{/pgf/pattern keys/line width}) and
    (\pgfkeysvalueof{/pgf/pattern keys/size},
0.5*\pgfkeysvalueof{/pgf/pattern keys/line width})},
  tile size={(\pgfkeysvalueof{/pgf/pattern keys/size},
\pgfkeysvalueof{/pgf/pattern keys/size})},
  tile transformation={rotate=\pgfkeysvalueof{/pgf/pattern keys/angle}},
  defaults={
    size/.initial=5pt,
    angle/.initial=45,
    line width/.initial=.4pt,
  },
  code={
      \draw [line width=\pgfkeysvalueof{/pgf/pattern keys/line width}]
        (0,0) -- (\pgfkeysvalueof{/pgf/pattern keys/size},0);
  },
}

\pgfdeclarepattern{
  name=hatch,
  parameters={\hatchsize,\hatchangle,\hatchlinewidth},
  bottom left={\pgfpoint{-.1pt}{-.1pt}},
  top right={\pgfpoint{\hatchsize+.1pt}{\hatchsize+.1pt}},
  tile size={\pgfpoint{\hatchsize}{\hatchsize}},
  tile transformation={\pgftransformrotate{\hatchangle}},
  code={
    \pgfsetlinewidth{\hatchlinewidth}
    \pgfpathmoveto{\pgfpoint{-.1pt}{-.1pt}}
    \pgfpathlineto{\pgfpoint{\hatchsize+.1pt}{\hatchsize+.1pt}}
    \pgfpathmoveto{\pgfpoint{-.1pt}{\hatchsize+.1pt}}
    \pgfpathlineto{\pgfpoint{\hatchsize+.1pt}{-.1pt}}
    \pgfusepath{stroke}
  }
}

\tikzset{
  hatch size/.store in=\hatchsize,
  hatch angle/.store in=\hatchangle,
  hatch line width/.store in=\hatchlinewidth,
  hatch size=5pt,
  hatch angle=0pt,
  hatch line width=.5pt,
}

\begin{figure}
  \centering
  \begin{minipage}{\textwidth}
    \scalebox{0.92}{
      \begin{tikzpicture}
      \centering
      \begin{axis}[
        height=4.5cm,
        width=1.1\linewidth,  
        bar width=0.33cm,  
        xmin=0, xmax=7,  
        axis x line*=bottom,
        axis y line*=left,
        ybar=6.4pt,
        enlarge x limits=0.08,  
        xtick={0, 1, 2, 3, 4, 5, 6, 7},
        xticklabel style={
            font=\scriptsize,
            yshift=-1mm,      
            text height=1.5ex,
            text depth=1ex,   
            align=center      
        },
        xticklabels={
            Quadcopter, Pendulum-v1, Pendulum-v2, Cart-Pole-v1,
            Cart-Pole-v2, Self-Driving-v1, Self-Driving-v2, 4-Car Platoon
        },
        ymin=0, ymax=100,
        ytick={0, 25, 50, 75, 100},
        yticklabels={0\%, 25\%, 50\%, 75\%, 100\%},
        yticklabel style={font=\normalsize},
        ymajorgrids,
        major grid style={draw=black!20},
        tick align=inside,
        tickwidth=1pt,
        y axis line style={opacity=0},
        legend style={
            at={(0.5, 1.1)}, anchor=south,
            legend columns=3, draw=none, fill=none, font=\normalsize
        },
        clip=false,
        every x tick label/.append style={
            rotate=0,          
            anchor=north,      
            inner sep=0pt      
        }
    ]

          \addplot [pattern={mylines[size=1.5pt,line width=1.1pt]}, pattern color=myred, draw=mydrawgray, line width=0.7pt] coordinates {
            (0, 66.554)
            (1, 82.00)
            (2, 71.43)
            (3, 38.536)
            (4, 50.33)
            (5, 21.768)
            (6, 39.09)
            (7, 77.77)
          };

          \addplot [pattern=hatch, pattern color=mygreen, hatch size=2pt, draw=mydrawgray, line width=0.7pt] coordinates {
            (0, 75.107)
            (1, 95.555)
            (2, 89.07)
            (3, 46.895)
            (4, 55.10)
            (5, 30.718)
            (6, 47.98)
            (7, 88.23)
          };
\node[above, font=\footnotesize] at (axis cs:0-0.19, 66.6) {66.6\%};
\node[above, font=\footnotesize] at (axis cs:1-0.19, 82.0) {82.0\%};
\node[above, font=\footnotesize] at (axis cs:2-0.19, 71.4) {71.4\%};
\node[above, font=\footnotesize] at (axis cs:3-0.19, 38.5) {38.5\%};
\node[above, font=\footnotesize] at (axis cs:4-0.19, 50.3) {50.3\%};
\node[above, font=\footnotesize] at (axis cs:5-0.19, 21.8) {21.8\%};
\node[above, font=\footnotesize] at (axis cs:6-0.19, 39.0) {39.0\%};
\node[above, font=\footnotesize] at (axis cs:7-0.19, 77.8) {77.8\%};

\node[above, font=\footnotesize] at (axis cs:0+0.19, 75.1) {75.1\%};
\node[above, font=\footnotesize] at (axis cs:1+0.19, 95.6) {95.6\%};
\node[above, font=\footnotesize] at (axis cs:2+0.19, 89.1) {89.1\%};
\node[above, font=\footnotesize] at (axis cs:3+0.19, 46.9) {46.9\%};
\node[above, font=\footnotesize] at (axis cs:4+0.19, 55.1) {55.1\%};
\node[above, font=\footnotesize] at (axis cs:5+0.19, 30.7) {30.7\%};
\node[above, font=\footnotesize] at (axis cs:6+0.19, 48.0) {48.0\%};
\node[above, font=\footnotesize] at (axis cs:7+0.19, 88.2) {88.2\%};

          \addlegendentry{\textsc{VRL} (avg. 55.9\%)\ \ \ \ \ \ }
          \addlegendentry{\tool (avg. 66.1\%)}
      \end{axis}
      \end{tikzpicture}
    }
  \end{minipage}
  \caption{The ratio of necessary interventions to total number of interventions for \tool and \textsc{VRL}.}
  \label{fig:RQ3}
\end{figure}

%% file: figures/RQ3.tex
\begin{figure}
    \begin{minipage}{0.75\textwidth}
      \centering
      \begin{tikzpicture}
      \centering
      \begin{axis}[
        height=4cm, width=\linewidth,
            axis x line*=bottom, axis y line*=left,
            tick align=inside,
            tickwidth=2.5pt,
            title style={at={(0.5,0)},anchor=north,yshift=-22},
            minor tick style={draw=none},
            ymin=0,  xticklabel style={yshift=-0.8mm, font=\normalsize, align=center}, yticklabel style={font=\normalsize, xshift=-0.8mm},
            x label style={{yshift=-0.3mm}},
            xmin=-0.2,
            xmax=2.2,
            xtick={0, 1, 2},
            xticklabels={Small (0.4 M), Medium (0.7 M), Large (7 M)},
            xlabel={\normalsize Scale of Neural Network},
            ymax=500,
            ytick={0, 125, 250, 375, 500},
            yticklabels={0s, 125s, 250s, 375s, 500s},
            ylabel={\normalsize Synthesis Time (s)},
            y label style={at={(0.2, 1.2)}, rotate=-90},
      ]
          \node[above] at ($(axis cs:1.5, 135)$) {\normalsize \tool};
          \node[above] at ($(axis cs:1.5, 415)$) {\normalsize \baseline};
          \addplot [draw=myred, mark=triangle, mark options={fill=myred}, mark size=2pt, line width=1.1pt] coordinates {
            (0, 28)
            (1, 300.2)
            (2, 441)
          };
          \addplot [draw=mygreen, mark=*, mark options={fill=mygreen}, mark size=2pt, line width=1.1pt] coordinates {
            (0, 14)
            (1, 79.2)
            (2, 169.5)
          };
    \end{axis}
    \end{tikzpicture}
\end{minipage}
\caption{Time required for the shield synthesis of neural policies at different sizes.}
\label{fig:RQ4}
\end{figure}

%% file: sections/discussion.tex
\section{Discussion}
\label{sec:discussion}

\subsection{Runtime Shields vs. Neural Policies}
\label{sec:discussion:runtime-shields-vs-neural-policies}

One question that remains unanswered is:
\begin{center}
    {\it Why do we need neural policies if runtime shields can control the systems?}
\end{center}
To answer this question, we compare the performance of neural policies and runtime shields in terms of the number of steps required to reach steady states. Figure~\ref{fig:steady-state} shows the results.

\input{figures/discussion.tex}

The figure shows that neural policies consistently require fewer steps to reach steady states compared to our runtime shields, with an average of 13.3 steps for neural policies versus 25.1 steps for runtime shields. This suggests that neural policies are more efficient at maintaining system stability. Therefore, we conclude that neural policies are essential for control systems, while runtime shields serve primarily to ensure the safety of these policies. Additionally, we observe that neural policies paired with \tool require more steps to reach steady states than those without shields, increasing the average steps from 13.3 to 23.6. This demonstrates that deploying runtime shields can compromise system stability, reflecting the ``no free lunch'' theorem~\cite{wolpert1997no}, where safety comes at the cost of stability. This observation somewhat aligns with the finding of Zhu et al.~\cite{zhu2019inductive}, which highlights that while alternative synthesis methods such as LQR-based approaches~\cite{10.7551/mitpress/8727.003.0004,reist2016feedback} can yield controllers with superior stability, they often fail to meet safety constraints. Our approach, therefore, represents a trade-off between safety and stability, but prioritizing safety. We leave the exploration of methods that can achieve better stability while maintaining safety to future work.

\subsection{Ablation Study}
\label{sec:discussion:ablation}

\begin{table}[t!]
    \renewcommand{\familydefault}{\sfdefault}\normalfont
    \centering
    \small
    \caption{Ablation study of \tool.}
    \label{tab:ablation}
    \scalebox{0.95}{
    \begin{tabular}{c|c|c|c|c}
        \hline\hline
                         & \# Safety Violations & Time Overhead & Memory Overhead & \# Shield Interventions \\ \hline\hline
    \tool &        0/1000 (100\% Safe)              &   1.28s     &       0.85 MB          &         313                \\\hline
     w/o Synthesis &        43/1000 (4.3\% Unsafe)              &   2.10s     &       0.88 MB          &         390                \\\hline
        w/o Optimization &        117/1000 (11.7\% Unsafe)              &   3.99s     &       0.85 MB          &         642                            \\ \hline\hline
\end{tabular}
    }
\end{table}

We perform an ablation study by comparing the performance of \tool with two variants, each lacking a key component. Table~\ref{tab:ablation} reports the average results across all eight control systems on the main evaluation metrics. Specifically, these two variants are:

\paragraph{w/o Synthesis}
This variant omits the counterexample-guided inductive synthesis process (Section~\ref{alg:counterexample-guided-synthesis}) and instead directly uses a Linear-Quadratic Regulator (LQR) to initialize a linear policy. This policy is then paired with the optimized switching strategy to construct the runtime shield. However, without counterexample-driven refinement, the policy is not guaranteed to be safe across the full state space, as it is not iteratively improved using concrete counterexamples. As shown in Table~\ref{tab:ablation}, this variant results in an average of 43 safety violations—substantially more than the 0 violations achieved by the full version of \tool. Although time and memory overheads and intervention counts are only moderately affected, this is expected: time overhead and interventions are primarily governed by the switching strategy, and memory usage is dominated by the linear policy structure. These findings underscore the importance of the synthesis process for ensuring robust safety guarantees.

\paragraph{w/o Optimization}
This variant removes the Bayesian optimization component (Section~\ref{sec:switching-strategy-synthesis}) used to synthesize the switching threshold. Instead, it employs randomly generated threshold values. As reported in Table~\ref{tab:ablation}, this results in an average of 117 safety violations—considerably more than in the full version of \tool. The time overhead also increases significantly from 1.28s to 3.99s, representing a 3.1$\times$ increase. Additionally, the number of shield interventions rises to 642, more than double the count in \tool. This suggests that the shield becomes both less effective and less efficient at preventing unsafe commands. The underlying issue is that non-optimized thresholds can either lead to over-aggressive interventions (causing unnecessary overhead) or too-lenient interventions (allowing unsafe commands to pass through). These results highlight that Bayesian optimization is essential not only for maximizing safety but also for improving runtime efficiency and minimizing unnecessary interventions.

\subsection{Fidelity of Inferred Environment Dynamics}
\label{sec:discussion:fidelity}

Our synthesis procedure assumes that the environment dynamics can be accurately inferred. To assess this assumption, we conduct experiments comparing the outputs of the inferred dynamics with those of the actual ones. Following a similar evaluation setup as in our RQs, we run each control system for 5,000 time steps using both the original and inferred dynamics. We collect the resulting state trajectories and quantify their difference using the mean squared error (MSE), defined as:
\begin{equation}
    \text{MSE} = \frac{1}{n} \sum_{i=1}^{n} (y_i - \hat{y}_i)^2
\end{equation}
where $y_i$ is the actual state at time step $i$, $\hat{y}_i$ is the predicted state from the inferred dynamics, and $n = 5{,}000$ is the number of samples. A lower MSE indicates higher fidelity between the inferred and actual dynamics.
The results are presented in the first column of Table~\ref{tab:mae}. In our experiments, we use the initial state of the system, which is assumed to be within a neighborhood of the equilibrium point, as the input for inferring environment dynamics. The results show an average MSE of only 3.2e-8, with a maximum MSE of 9.9e-8, both of which are extremely low. These small error values suggest that the inferred dynamics closely approximate the true system dynamics, indicating that any adverse effects on the shield synthesis process are likely negligible.

\begin{table}[t!]
    \centering
    \caption{The mean absolute error (MAE) between the inferred and actual environment dynamics.}
    \label{tab:mae}
    \renewcommand{\familydefault}{\sfdefault}\normalfont
    \scalebox{0.95}{
    \begin{tabular}{c|c|c}
    \hline\hline
    \multirow{2}{*}{Control System} & \multicolumn{2}{c}{MAE}\\ \cline{2-3}
                                             & random initial state & equilibrium point \\ \hline\hline
    Quadcopter         & 1.4e-7  & 6.2e-10 \\ \hline
    Pendulum-v1        & 3.9e-10 & 1.1e-13 \\ \hline
    Pendulum-v2        & 5.6e-9  & 0.0     \\ \hline
    CartPole-v1        & 4.3e-10 & 0.0     \\ \hline
    CartPole-v2        & 2.2e-11 & 2.3e-11 \\ \hline
    Self-Driving-v1    & 2.5e-9  & 1.8e-9  \\ \hline
    Self-Driving-v2    & 6.3e-9  & 1.8e-9  \\ \hline
    4-Car Platoon      & 9.9e-8  & 0.0     \\ \hline\hline
    Average            & 3.2e-8  & 5.3e-10 \\ \hline
    \end{tabular}
    }
\end{table}

\paragraph{Impact of Equilibrium Points}
In our experiments, for each run, we randomly sample the initial state of the system over the set of initial states. This approach ensures that the evaluation is not biased toward any specific starting condition. However, as described in Section~\ref{sec:state-transition}, environment dynamics inference tends to be most accurate at equilibrium points where the system is stable and behaves in a locally linear and time-invariant manner. To assess the impact of using equilibrium points, we conduct a supplementary experiment in which we manually compute the equilibrium point for each control system and use it as the initial state for dynamics inference. The results, shown in the second column of Table~\ref{tab:mae}, indicate a further reduction in MSE, with an average value of just 5.3e-10. While this confirms that inference at equilibrium points yields even higher fidelity, the improvement is minimal in absolute terms, with differences appearing only at the eighth decimal place. This suggests that, in our setting, the shield synthesis procedure remains robust even when dynamics are inferred from arbitrary (non-equilibrium) initial states. Since equilibrium points may not be known in advance in practical scenarios, we believe that using randomly sampled initial states for inferring environment dynamics remains a reasonable and effective choice.

\subsection{Limitations of \tool}
\label{sec:discussion:limitations}

While the above evaluation demonstrates the effectiveness of \tool in synthesizing runtime shields for neural policies across diverse control systems, we acknowledge a key limitation: \tool assumes that safety specifications define convex regions, typically expressed as conjunctions of linear inequalities. This limitation stems from the mathematical foundation of \tool, which relies on linear programming to compute the maximal output admissible set (see Section~\ref{sec:counterexample-guided-synthesis}). Disjunctive specifications, which define disconnected or hole-containing regions in the state space, are non-convex and thus fall outside the scope of linear programming-based safety verification. Importantly, this convexity assumption is not unique to \tool. Prior methods such as VRL~\cite{zhu2019inductive} make the same assumption, as their synthesis procedure constructs barrier certificates whose safe regions must be convex to enable tractable verification using convex optimization techniques such as sum-of-squares programming.

Despite this theoretical limitation, we argue that many real-world safety constraints—such as bounds on velocity, acceleration, or joint angles—are naturally convex and can be effectively represented by conjunctions of linear inequalities. Therefore, while extending support to non-convex safety specifications remains a promising direction for future work, the current formulation of \tool already addresses a wide range of practical and safety-critical scenarios and thus remains valuable.

\subsection{Threats to Validity}

One threat to internal validity is the random behavior of the optimization algorithms used in our experiments. To mitigate this threat, we repeat each experiment 10 times and use appropriate statistical tests to account for both statistical significance and effect size, as recommended by Arcuri and Briand~\cite{arcuri2011practical}. Therefore, we believe that the randomness of the results is well-controlled. Another potential threat to internal validity is the implementation of \tool and our scripts. To reduce this threat, we have carefully reviewed the code before submission and made it publicly available for review and improvement.

In terms of external validity, one potential threat is that the results of our analysis may not be fully generalizable. To mitigate this, we carefully selected a representative set of control systems with diverse tasks to evaluate our method and baseline under various conditions, ensuring that our results are not biased. However, it is important to note that our method relies on the assumption that the system dynamics are sufficiently linearizable when inferring environment dynamics (see Section~\ref{sec:state-transition}), which may not be true for all control systems. Consequently, even though the control systems used in our experiments are non-linear, the generalizability of our method to a broader class of nonlinear systems warrants further investigation in future work.

One potential threat to construct validity is that the evaluation metrics may not fully capture the performance of falsification tools. To mitigate it, we used a total of six commonly-used evaluation metrics that are introduced in Section~\ref{sec:Experimental_Setup} to compare the performance of \tool and the baseline from a comprehensive set of perspectives.

%% file: figures/discussion.tex
\begin{figure}
    \centering
   \begin{minipage}{\textwidth}
   \scalebox{0.92}{
   \begin{tikzpicture}
   \centering
   \pgfplotsset{
       mystyle/.style={
           nodes near coords,
           nodes near coords style={font=\normalsize, text=black, anchor=south},
           point meta=explicit symbolic
       }
   }
   \begin{axis}[
    height=4.0cm, width=1.1\linewidth,
    /pgf/bar width=0.28cm,
    xmin=-0.4, xmax=7.4,
    axis x line*=bottom, axis y line*=left, enlarge x limits=true,
    enlarge x limits=0.02, 
    xtick={0, 1, 2, 3, 4, 5, 6, 7},
    xticklabel style={
    font=\scriptsize,
    yshift=-1mm, 
    text height=1.5ex,
    text depth=1ex, 
    align=center 
    },
    xticklabels={
    Quadcopter, Pendulum-v1, Pendulum-v2, Cart-Pole-v1,
    Cart-Pole-v2, Self-Driving-v1, Self-Driving-v2, 4-Car Platoon
    },
    ybar=3.8pt, clip=false,
    ymin=0, ymax=60, ytick={0, 15, 30, 45, 60}, yticklabels={0, 15, 30, 45, 60},
    ymajorgrids, major grid style={draw=black!20}, tick align=inside,
    yticklabel style={font=\normalsize}, tickwidth=0pt,
    y axis line style={opacity=0},
    legend style={at={(0.5,1.1)}, anchor=south, legend columns=3, draw=none, fill=none, font=\normalsize},
    every node near coord/.append style={text=black} 
    ]

   \addplot+[mystyle, pattern=hatch, pattern color=mygreen, hatch size=2pt, draw=mydrawgray, line width=0.7pt] coordinates {
    (0, 22) [22]
    (1, 4) [4]
    (2, 3) [3]
    (3, 7) [7]
    (4, 17) [17]
    (5, 9) [9]
    (6, 14) [14]
    (7, 30) [30]
    };

   \addplot+[mystyle, pattern={mylines[size=1.5pt,line width=1.1pt,angle=-40]},
    pattern color=gray, draw=mydrawgray, line width=0.7pt] coordinates {
    (0, 24) [24]
    (1, 15) [15]
    (2, 8) [8]
    (3, 25) [25]
    (4, 21) [21]
    (5, 18) [18]
    (6, 31) [31]
    (7, 59) [59]
    };

   \addplot+[mystyle, pattern={mylines[size=1.5pt,line width=1.1pt]}, pattern color=myred, draw=mydrawgray, line width=0.7pt] coordinates {
    (0, 28) [28]
    (1, 8) [8]
    (2, 7) [7]
    (3, 16) [16]
    (4, 19) [19]
    (5, 18) [18]
    (6, 33) [33]
    (7, 60) [60]
    };

   \addlegendentry{Neural Policy (avg. 13.3) \ \ \ \ }
   \addlegendentry{\textsc{Aegis} (avg. 25.1) \ \ \ \ }
   \addlegendentry{Neural Policy w/ \textsc{Aegis} (avg. 23.6) }
   \end{axis}
   \end{tikzpicture}
    }
   \end{minipage}
   \caption{Time steps to reach steady states.}
   \label{fig:steady-state}
   \end{figure}

%% file: sections/related.tex
\section{Related Work}
\label{sec:related}

This section reviews the related literature and positions our work in the context of existing research.

\paragraph{Runtime Shields for Neural Policies}
Runtime shields have been used in runtime enforcement for reactive systems~\cite{wu2019shield,konighofer2017shield,akametalu2014reachability,gillula2012guaranteed,bloem2015shield,konighofer2020shield,alshiekh2018safe,yang2021biasheal} long before the advent of neural policies. However, runtime shields for neural policies are a relatively new concept and have only been studied in a few papers. Runtime shields in reinforcement learning models were first introduced in~\cite{alshiekh2018safe}. Because their verification approach is not symbolic, however, it can only work over finite discrete states instead of continuous state settings like the ones we consider in this paper. Zhu et al.~\cite{zhu2019inductive} proposed a method to synthesize deterministic and verified programs from neural policies, which can be used as runtime shields for neural policies. However, their shields have limitations in terms of efficiency and permissiveness, which necessitates the need for a more effective and permissive approach, motivating our work. Xiong et al.~\cite{xiong2021scalable} propose a method capable of synthesizing high-quality runtime shields even when the subject system involves hundreds of dimensions. RMPS~\cite{9196867} achieves runtime shielding via backup model-predictive control (MPC) controllers that can override unsafe commands from neural policies. However, both methods are not sufficiently effective to synthesize runtime shields that can guarantee safety but still occasionally let unsafe commands pass through. Therefore, we do not consider it as a baseline in our experiments. 

Several other works aim to enhance the safety of neural policies during training by incorporating runtime shield feedback into the learning loop. For instance, REVEL~\cite{NEURIPS2020_448d5eda} uses shield interventions as feedback to iteratively refine the learned policy via imitation learning. Similarly, VEL~\cite{10.1007/978-3-031-30820-8_16} leverages verification feedback to iteratively optimize programmatic controllers with provable safety guarantees. In addition to optimization via runtime shield feedback, other approaches improve policy safety through specialized policy optimization techniques, such as policy gradient~\cite{papini2022smoothing}. For example, Yu et al.~\cite{NEURIPS2019_db29450c} propose a convergent policy gradient algorithm for safety-constrained reinforcement learning, while Honari et al.~\cite{10611316} introduce a multi-objective optimization approach that balances reward and safety. Verma et al.~\cite{NEURIPS2019_5a44a53b} further combine policy gradient methods with program synthesis to learn safe and verifiable policies, with theoretical convergence guarantees. Moreover, a separate line of work considers formal verification and synthesis of neural policies without relying on runtime shielding. This includes methods that apply star set-based reachability analysis~\cite{10.1145/3358230,10383593}, leverage Lyapunov functions~\cite{icse2025,NEURIPS2018_4fe51490,NEURIPS2019_2647c1db,abate2020formal}, or utilize supermartingales~\cite{ansaripour2023learning,chatterjee2023learner,vzikelic2023learning,9990576} to provide rigorous safety guarantees.

\paragraph{Program Synthesis for Neural Policies}
Although program synthesis has been studied for decades~\cite{sobania2022comprehensive}, very little attention has been paid to program synthesis specialized for neural policies. A few studies~\cite{verma2018programmatically,trivedi2021learning} have proposed methods for generating programmatic policies from well-trained DRL models, which are more interpretable than neural networks. Additionally, Inala et al.~\cite{inala2020synthesizing} and Qiu et al.~\cite{qiu2021programmatic} synthesize programmatic controllers to perform continuous control tasks. Moreover, there are a few studies that focus on synthesizing code implementations or configurations for deep learning models in various tasks~\cite{10.1145/3650212.3652119,10.1145/3639475.3640097,285489,10.1145/3551349.3556964,10.1145/3708525}. Our approach differs from these studies in that we propose a counterexample-guided inductive program synthesis approach which can efficiently synthesize runtime shields for neural policies, which is a unique application of program synthesis for neural policies.

Another line of work~\cite{mandal2023,10.1145/3377930.3390213,NEURIPS2018_7aa685b3,pmlr-v70-gaunt17a} formulates program synthesis as a continuous optimization problem, rather than selecting discrete grammar components as in many syntax-guided approaches~\cite{balog2017deepcoder,bunel2018leveraging}. While our method shares this high-level formulation, i.e., seeking optimal parameters in a continuous space, it differs significantly in both objective and technical design. Concretely, we focus on synthesizing runtime shields guided by formal safety constraints, whereas prior work targets functional correctness from input-output examples~\cite{mandal2023,10.1145/3453483.3454080}, natural language specifications~\cite{mandal2023largelanguagemodelsbased,10.1145/2884781.2884786}, or even visual inputs such as images~\cite{10.1145/3528233.3530715,NEURIPS2018_7aa685b3}. Meanwhile, our approach technically diverges by using refinement based on the maximal admissible set and applying Bayesian optimization, distinguishing it from prior methods that rely on evolutionary strategies~\cite{mandal2023,10.1145/3377930.3390213} or neural networks~\cite{NEURIPS2018_7aa685b3,10.1145/3453483.3454080}.

\paragraph{Formal Verification of Neural Policies}
Formal verification of neural policies has been explored in several studies~\cite{landers2023deep, mazouni2023review}. Some of these studies align with our work by using counterexample-guided inductive synthesis (CEGIS) to formally verify neural policies and their neural certificates. Certain works, including~\cite{chang2019neural, abate2020formal, abate2021fossil, edwards2024fossil}, focus on deterministic systems, employing symbolic reasoning to verify neural policies and construct Lyapunov functions~\cite{haddad2008nonlinear} or barrier certificates in a CEGIS loop. Another more recent line of work~\cite{lechner2022stability, vzikelic2023learning, chatterjee2023learner,ansaripour2023learning} considers both deterministic and stochastic systems, utilizing discretization and abstract interpretation for verification within the CEGIS loop, and deriving neural certificates of correctness or stability in the form of supermartingales instead of Lyapunov and barrier functions. Our work, however, differs by focusing on synthesizing runtime shields as simple programs. This approach aims to enhance the efficiency and permissiveness of these shields, leading to a unique application of formal verification in the context of neural policies.


\paragraph{Linear Constraints of Neural Networks}
There is a rich body of work on designing neural networks whose outputs or activations satisfy a set of linear constraints by construction~\cite{6030948,hendriks2021linearlyconstrainedneuralnetworks,Frerix_2020_CVPR_Workshops,Pathak_2015_ICCV,stoian2024how,tordesillas2023rayenimpositionhardconvex}. Among these, Frerix et al.~\cite{Frerix_2020_CVPR_Workshops} enforce hard convex constraints on the activations of deep neural networks through a reparameterization of the feasible set. Hendriks et al.~\cite{hendriks2021linearlyconstrainedneuralnetworks} and Tordesillas et al.~\cite{tordesillas2023rayenimpositionhardconvex} focus on imposing linear constraints on neural network outputs, even under low-data conditions or with very low inference latency. Stoian et al.~\cite{stoian2024how} introduce a constraint layer for generative models that guarantees user-defined linear constraints on tabular data, ensuring that generated samples remain realistic and domain-compliant. In addition, Giunchiglia et al.~\cite{giunchiglia2023road,doi:10.3233/NAI-240767} recently propose methods and a dataset for training networks that adhere to safety requirements expressed as linear or logical constraints. While these approaches enforce constraints that may be interpreted as safety properties, they assume the network can be retrained and its internal parameters are accessible. In contrast, as discussed in Section~\ref{sec:intro}, our work addresses a fundamentally different setting when retraining the neural policy is prohibitively expensive, or its internal structure is inaccessible, such as in proprietary third-party systems.

%% file: sections/conclusion.tex
\section{Conclusion and Future Work}
\label{sec:conclusion}

In this paper, we present \tool, a novel approach to synthesizing lightweight and permissive programmatic runtime shields for neural policies. \tool achieves this by formulating the seeking of a runtime shield as a sketch-based program synthesis problem and using two novel algorithms to solve it, namely a counterexample-guided inductive synthesis algorithm and a Bayesian optimization algorithm. To evaluate \tool and its synthesized shields, we use eight representative control systems with neural policies and compare \tool to the current state-of-the-art \baseline. Our results show that the programmatic runtime shields synthesized by \tool can correct all unsafe commands from neural policies, ensuring that the systems do not violate the desired safety properties at any time. In particular, compared to \baseline, \tool's runtime shields exhibit a 2.2$\times$ reduction in time overhead and a 3.9$\times$ reduction in memory overhead, suggesting that they are much more lightweight. Additionally, \tool's shields show improved permissiveness, leading to an average of 1.5$\times$ fewer interventions compared to \baseline's shields. These results not only highlight the superiority of \tool over the state-of-the-art, but also show that \tool is an effective framework for synthesizing programmatic runtime shields for neural policies.

In the future, we plan to extend \tool to support more complex control systems that process richer inputs, such as images~\cite{yang2022revisiting} and audio~\cite{10.1145/3707450}. We also intend to develop more advanced program synthesis techniques to further improve \tool, for example by incorporating neural-guided synthesis~\cite{10.1145/3763102} or Large Language Models~\cite{10.1145/3695988,10.1145/3731753,9825794} to enhance the efficiency of the synthesis process.

\ans{
\textbf{Replication Package.}\hspace{6pt} For reproducibility and advancing the state of research, we have made our tool and all experimental data publicly available: \url{https://github.com/soarsmu/AEGIS}.}